\documentclass[structabstract]{aa} 
\usepackage{amsmath}
\usepackage{comment}
\usepackage{graphicx}
\usepackage{txfonts}
\usepackage{natbib}
\usepackage{longtable}
\usepackage{lscape}
\usepackage{multirow}
\usepackage{pdfpages}

\usepackage[colorlinks=true, allcolors=blue]{hyperref}

\bibpunct{(}{)}{;}{a}{}{,}

\newcommand{\kms}{km\,s$^{-1}$}

\newcommand{\degree}{$^{\circ}$}

\begin{document}

\title{A global view on star formation: The GLOSTAR Galactic plane survey}
\subtitle{XII. Effelsberg's continuum view and data release}
\author{Y.~Gong\inst{1,2}, W.~Reich\inst{1}, M.~R.~Rugel\inst{3,4,1}\fnmsep\thanks{Jansky Fellow of the National Radio Astronomy Observatory.}, K.~M.~Menten\inst{1}\fnmsep\thanks{During the preparation of this article, we suffered the tragic loss of Prof. Karl M. Menten, the principal investigator of the GLOSTAR survey. We dedicate this work to his enduring legacy. Karl’s insight and vision shaped not only this project but also the field of radio astronomy at large. We are deeply indebted to the countless discussions with him that have inspired generations of radio astronomers. We will always miss his warmth, wisdom, boundless curiosity, generosity of spirit, kindness, and delightful sense of humor.}, A.~Brunthaler\inst{1}, F.~Wyrowski\inst{1},  P.~M{\"u}ller\inst{1}, S.~A.~Dzib\inst{1}, J.~S.~Urquhart\inst{5}, A.~Y.~Yang\inst{6,7,1}, R.~Dokara\inst{1}, G.~N.~Ortiz-León\inst{8}, B.~Winkel\inst{1}, A.~Kraus\inst{1}, S.~P.~Sathyanarayanan\inst{1}, W.-J.~Kim\inst{9}, H.~Beuther\inst{10}, J.~D.~Pandian\inst{11}, A.~Cheema\inst{1}, S~Khan\inst{1}, V.~S.~Veena\inst{1}, N.~Roy\inst{12}, C.~Carrasco-Gonzalez\inst{1}, W.~Cotton\inst{3}, T.~Csengeri\inst{13}, S.-N.~X. Medina\inst{1}, H.~Nguyen\inst{1}}

\institute{
Max-Planck-Institut f{\"u}r Radioastronomie, Auf dem H{\"u}gel 69, D-53121 Bonn, Germany
\and 
Purple Mountain Observatory, Chinese Academy of Sciences, 10 Yuanhua Road, Nanjing 210023, PR China
\and
National Radio Astronomy Observatory, PO Box O, Socorro, NM 87801, USA
\and 
Center for Astrophysics \texttt{|} Harvard \& Smithsonian, 60 Garden Street, Cambridge, MA 02138, USA
\and
Centre for Astrophysics and Planetary Science, University of Kent, Canterbury CT2 7NH, UK
\and 
National Astronomical Observatories, Chinese Academy of Sciences, Beijing 100101, China
\and 
Key Laboratory of Radio Astronomy and Technology, Chinese Academy of Sciences, A20 Datun Road, Chaoyang District, Beijing, 100101, PR China
\and 
Instituto Nacional de Astrofísica, Óptica y Electrónica, Apartado Postal 51 y 216, 72000 Puebla, México
\and 
I. Physikalisches Institut, Universit{\"a}t zu K{\"o}ln, Z{\"u}lpicher Str. 77, 50937 K{\"o}ln, Germany
\and 
Max Planck Institute for Astronomy, K{\"o}nigstuhl 17, 69117 Heidelberg, Germany
\and 
Department of Earth and Space Science, Indian Institute for Space Science and Technology, Trivandrum 695547, India
\and 
Department of Physics, Indian Institute of Science, Bengaluru, 560012, India
\and 
Laboratoire d’astrophysique de Bordeaux, Univ. Bordeaux, CNRS, B18N, allée Geoffroy Saint-Hilaire, 33615 Pessac, France
}

\date{Received date ; accepted date}

\abstract
{Extended radio continuum emission and its linear polarization play a key role in probing large-scale structures of synchrotron and free-free emission in the Milky Way. Despite the existence of many radio continuum surveys, sensitive and high-angular-resolution single-dish surveys of extended radio continuum emission remain scarce.}
{Our objective is to deliver a Galactic plane survey of extended radio continuum emission within the frequency range of 4--8~GHz, achieving an unprecedented angular resolution of $\lesssim$3\arcmin. As part of the GLObal view of STAR formation (GLOSTAR) survey, we will also crucially complement the existing data from the Karl G. Jansky Very Large Array (VLA) by addressing the missing zero-spacing gap.}
{Within the framework of the GLOSTAR Galactic plane survey, we performed large-scale radio continuum imaging observations toward the Galactic plane in the range $-2^{\circ}< \ell <60^{\circ}$ and $|b|<1.1^{\circ}$, as well as the Cygnus X region $76^{\circ}< \ell <83^{\circ}$ and $-1^{\circ}<b<2^{\circ}$ with the Effelsberg 100-m radio telescope.}
{We present the Effelsberg continuum survey at 4.89~GHz and 6.82~GHz including linear polarization with angular resolutions of 145\arcsec\, and 106\arcsec, respectively. The survey has been corrected for missing large-scale emission using available low-angular-resolution surveys. Comparison with previous single-dish surveys indicates that our continuum survey represents the highest-quality single-dish data collected to date at this frequency. More than 90\% of the flux density missed by the VLA D-array data is effectively recovered by the Effelsberg continuum survey. The improved sensitivity and angular resolution of our survey enable reliable mapping of Galactic magnetic field structures, with polarization data that are less affected by depolarization than in previous surveys. The GLOSTAR single-dish continuum data will be released publicly, offering a valuable resource for studying extended objects including H{\scriptsize II} regions, supernova remnants, diffuse interstellar medium, and Galactic structure.}
{}


\titlerunning{Effelsberg's insight into the Galactic plane's radio continuum emission}

\authorrunning{Gong et al.}

\maketitle

\section{Introduction}
Ionized gas constitutes a significant component of the interstellar medium (ISM), accounting for $\sim$20\% of the total gas mass within the Milky Way \citep[e.g.,][]{2010pcim.book.....T,2011piim.book.....D}. This component fills the gaps between the colder and denser phases of the ISM, namely the cold and warm neutral media (CNM and WNM). Widespread throughout the Galactic plane, ionized gas exerts a critical influence on the Galaxy’s dynamics, energy balance, and overall structure. Therefore, investigating the properties of the Galactic ionized gas is essential for a comprehensive understanding of the physical processes of star formation, H{\scriptsize II} regions, supernova remnants (SNRs), stellar feedback, and Galactic structures.

Radio continuum surveys are indispensable for studying ionized gas in the Milky Way, detecting both nonthermal synchrotron and thermal bremsstrahlung (i.e., free-free) radiation from electrons in ionized gas \citep[e.g.,][]{2013tra..book.....W,2016era..book.....C}. These emissions, prominent at centimeter wavelengths, can penetrate dust clouds that obscure observations at optical wavelengths, thereby providing an unobstructed view of the ionized ISM.

A large number of single-dish and interferometric radio continuum surveys have already been carried out to study the Galactic plane \citep[e.g.,][]{1984A&AS...58..197R,1998AJ....115.1693C,2016A&A...595A..32B,2023MNRAS.524.1291P}. 
While interferometric radio continuum observations provide a high-angular-resolution view of the radio continuum emission, 
the missing short spacings in the \textit{uv}-space make these observations insensitive to large-scale radio continuum structures \citep[see Fig.~1 in][]{2019A&A...627A.175M}. Combining single-dish and interferometric data enables a Galactic plane survey to capture continuum and polarized emission structures across all angular scales larger than the interferometer's angular resolution \citep{Landecker10,2020A&A...634A..83W,2021A&A...651A..85B,2023A&A...671A.145D}. Hence, single-dish radio continuum surveys are important to recover the large-scale radio continuum structure. 

Single-dish radio continuum studies of the Milky Way date back to the pioneering work of \citet{1944ApJ...100..279R}. Existing radio continuum Galactic plane surveys with angular resolutions of 20\arcmin\ or better
conducted with ground-based single-dish telescopes span frequencies from around 1~GHz up to $\sim$10~GHz
\citep{Altenhoff70,1989AJ.....97.1064C, Reich90,Fuerst90, 1990A&AS...83..539R, 
1987PASJ...39..709H,2011A&A...527A..74S,2015MNRAS.448.3572I,2025arXiv250114203S}, while space telescopes such as the Wilkinson Microwave Anisotropy Probe (\textit{WMAP}) and the \textit{Planck} space telescope cover higher frequencies than $\sim$23~GHz \citep{2007ApJS..170..335P,2020A&A...641A...1P}. However, all single-dish radio continuum surveys have angular resolutions of $\gtrsim$2\rlap{.}\arcmin7, indicating a scarcity of high-angular-resolution single-dish radio continuum surveys. Our Effelsberg radio continuum survey aims to address this gap, achieving an angular resolution of $\lesssim$2\rlap{.}\arcmin4. While the angular resolution of our survey is similar to the early Effelsberg 5~GHz survey by \citet{Altenhoff79}, the survey sensitivity will be significantly enhanced, thanks to advancements in receiver system technology.

As part of the Global View on Star Formation in the Milky Way (GLOSTAR\footnote{\url{https://glostar.mpifr-bonn.mpg.de/glostar/}}) survey \citep{2019A&A...627A.175M,2021A&A...651A..85B}, we conducted the large-scale radio continuum imaging observations of the Galactic plane at 4--8~GHz with the Effelsberg 100-m radio telescope. These observations complement the Karl G. Jansky Very Large Array (VLA) survey, enhancing our study of Galactic structures on all scales down to $\sim$1\arcsec. The GLOSTAR data have already been used in a wide range of studies. These include the compilation of catalogs of compact radio continuum sources \citep{2019A&A...627A.175M,2023A&A...670A...9D,2023A&A...680A..92Y,2024A&A...689A.196M} and the estimation of the present-day star formation rate in the Central Molecular Zone \citep{2021A&A...651A..88N}. The data have also been employed to study radio recombination line (RRL) emission from bright H{\scriptsize II} regions \citep{2024A&A...689A..81K}, to identify the youngest H{\scriptsize II} regions and examine their variability \citep{2025A&A...694A..26Y}, and to characterize supernova remnants (SNRs) \citep{2021A&A...651A..86D,2023A&A...671A.145D}. In addition, GLOSTAR has enabled searches for 6.7~GHz class II CH$_3$OH masers \citep{2021A&A...651A..87O,2022A&A...666A..59N} and investigations of 4.8~GHz formaldehyde absorption in Cygnus X \citep{2023A&A...678A.130G}. The Effelsberg radio continuum component of the GLOSTAR survey will allow us to explore the large-scale radio continuum structures in the Milky Way. 

While smaller portions of our Effelsberg radio continuum observations, in combination with the VLA observations, have been previously reported in the earlier studies \citep[e.g.,][]{2021A&A...651A..85B,2023A&A...671A.145D,2023A&A...678A.130G}, these studies are limited to specific regions. Here, we extend this research by investigating the properties of radio continuum emissions in the Galactic plane, providing insights into the characteristics on a Galactic scale.

The observations and data reduction are described in Sect.~\ref{Sec:obs}. Subsequently, we present the observational results and discussions in Sect.~\ref{Sec:res}. Our findings are summarized in Sect.~\ref{Sec:sum}.

\section{Observations and data reduction}\label{Sec:obs}
\subsection{Observations with the Effelsberg 100 m telescope}\label{Sec:eff}
As an important component of the GLOSTAR survey \citep{2021A&A...651A..85B}, we performed large-scale 4--8~GHz radio continuum observations using the Effelsberg 100-m radio telescope\footnote{The 100-m telescope at Effelsberg is operated by the Max-Planck-Institut f{\"u}r Radioastronomie (MPIfR) on behalf of the Max-Planck-Gesellschaft (MPG).}. The observed regions include the Galactic plane in the range $-2^{\circ}< \ell <60^{\circ}$ and $|b|<1.1^{\circ}$, as well as the Cygnus X region ($76^{\circ}<\ell<83^{\circ}$ and $-1^{\circ}<b<2^{\circ}$). These observations were carried out between 2019 January 11 and 2023 August 23 (project codes $102-20$ and $22-15$). Our observations have a sky coverage of $\sim$145 square degrees in total. Although the continuum and spectral-line observations were conducted simultaneously, here we focus on the continuum and polarization observations. 
For the continuum observations, the SPEctro-POLarimeter (SPECPOL) backend was used to record full Stokes continuum emission in the MBFITS format\footnote{\url{https://fits.gsfc.nasa.gov/registry/mbfits.html.}}. SPECPOL offers two frequency bands, spanning 4-6 GHz (lower band) and 6-8 GHz (upper band), respectively. Each band is divided into 1024 channels, yielding a channel width of 1.95~MHz.

For observations, the Galactic plane in the range $-2^{\circ}<\ell<60^{\circ}$ and $|b|<1.1^{\circ}$ was split into 31 fields\footnote{See \url{http://gongyan2444.github.io/progress.html} for instance.}, while the Cygnus X region was split into three fields. In the pilot region ($28^{\circ}<\ell<36^{\circ}$ and $|b|<1^{\circ}$), each field is divided into rectangular cells of 0.2\degree$\times$2\degree\, \citep{2021A&A...651A..85B}. In Cygnus X, each field is divided into 0.1\degree$\times$3\degree\, cells. The rest of the fields were divided into rectangular cells of 0.2\degree$\times$2.2\degree. These cells facilitate flexible observing schedules.
Before the observations start, the high-speed analog-to-digital converters (ADC) are calibrated to minimize the offset, gain and phase (OGP) variations within each converter in a procedure similar to \citet{2014JAI.....350001P}. Following the OGP corrections, the time delay between the two signal paths are measured by computing the correlation of a common noise source that is fed to the ADCs. Subsequently this delay is compensated in the SPECPOL firmware, by delaying one of the signals appropriately before the Stokes parameters are computed.
The on-the-fly (OTF) mode was used to map each cell with a scanning speed of 90\arcsec\,per second and a step of 30\arcsec\,to fulfill the Nyquist sampling condition. Each region was mapped in both Galactic longitude and latitude in order to allow for using the basket weaving method to suppress the scanning artifacts \citep[e.g.,][]{2017A&A...606A..41M}. During the observations, the system temperatures were typically from 28~K to 42~K. Focus was verified after sunrise and sunset. Nearby pointing observations were carried out every 2 to 3 hours. The rms pointing uncertainty was found to be within 10\arcsec.

The calibrators 3C~286 and NGC~7027 were observed to derive accurate bandpass solutions and to compute the calibration fit for the SPECPOL data \citep{1994A&A...284..331O,2021A&A...651A..85B,2013ApJS..204...19P,2017ApJS..230....7P}. The calibration of the Stokes parameters ($I$, $U$, $Q$) was performed by applying the M{\"u}ller Matrix \citep[e.g.,][]{2001PASP..113.1274H} that was obtained by observing 3C~286 at different parallactic angles. 
To minimize radio-frequency interference (RFI) in the raw continuum data, we select discontinuous channels centered at 4.89~GHz and 6.82~GHz to produce the radio continuum images (see Fig.~\ref{Fig:spmask}), respectively. These channels cover a bandwidth of $\sim$120~MHz at both images. The narrow bandwidth also help to reduce the depolarization effects. According to maps of the unpolarized source 3C~295 (see Fig.~\ref{Fig:3c295}), the `on-axis' instrumental polarization is up to 0.5\% for both continuum bands, but higher within the ``butterfly-shaped" response within the main-beam area. This calibration was further validated by mapping 3C~286 at various parallactic angles. The resulting polarization angle agrees within 1\degree\ uncertainty, and the polarization fraction is consistent within a 1\% uncertainty, confirming the robustness of our calibration. Based on Gaussian fitting of unresolved continuum sources, we measured that the half-power beam widths (HPBWs) are 145\arcsec\,and 106\arcsec\,at 4.89~GHz and 6.82~GHz, respectively. The observations for the region at $22^{\circ}<\ell<26^{\circ}$ were significantly affected by geostationary satellites. To minimize the impact of RFI during satellite passages, this region was observed multiple times, and only apparently `clean' subscans were selected for subsequent analysis. The radio continuum data were reduced with the toolbox program and the NOD3 software package \citep{2017A&A...606A..41M}. 

Since the radio continuum emission is prevalent throughout the observed region, the rms noise levels are difficult to directly determine from the Stokes $I$ map. We made use of the constrained diffusion decomposition (CDD) method\footnote{The code is available from \url{https://github.com/gxli/Constrained-Diffusion-Decomposition}.} to decompose the Stokes $I$ map \citep{2022ApJS..259...59L}. We estimate the noise from the most compact component where extended emission is effectively removed, leading to there being only a few compact sources. We applied the statistics of the emission-free area to estimate the rms noise in each field, with the derived noise distribution presented in Fig.~\ref{Fig:noise}. The noise levels are within the range of 3.9--11.5~mK (i.e., 1.6--4.7~mJy~beam$^{-1}$) for the lower band and 3.2--8.9~mK (i.e., 1.4--3.8~mJy~beam$^{-1}$) for the upper band, respectively. The distribution is similar for the two bands, with slightly higher noises observed in the lower band. For the polarized emission, we see a similar decrease of the noise with increasing Galactic longitude with a median value of about 7~mK for both frequencies. 
Additionally, elevated noise levels were noted in fields near the Galactic center, attributed to observations conducted at low elevations. The observational parameters of this continuum survey are summarized in Table~\ref{Tab:obs}.

\begin{table}[!hbt]
\caption{Observational parameters of the Effelsberg continuum survey.}\label{Tab:obs}
\normalsize
\centering
\begin{tabular}{ccc}
\hline \hline
Parameters  & lower band &  upper band    \\ 
\hline
Central frequency  &  4.89~GHz         &   6.82~GHz             \\
System temperature & 28--42~K           & 28--42~K                 \\
HPBW               & 145\arcsec        &   106\arcsec           \\
Bandwidth          & $\sim$120~MHz     &   $\sim$120~MHz        \\
Grid               & 30\arcsec$\times$30\arcsec  & 30\arcsec$\times$30\arcsec \\
1$\sigma$ rms for I& 3.9--11.5~mK     &   3.2--8.9~mK        \\
1$\sigma$ rms for Polarized Intensity& 5.8--13.0~mK     &   5.6--13.8~mK        \\
Conversion factor ($T_{\rm B}$/$S_{\nu}$) & 2.436~K/Jy & 2.34~K/Jy\\
\hline
\end{tabular}
\normalsize
\end{table}

\subsection{Zero-level restoration}\label{sec.zero}
Due to the limited latitude coverage of our observations, the zero level is not reached, and the reduced radio continuum images have offsets from the absolute zero level. Similar to our previous efforts \citep{2021A&A...651A..85B,2023A&A...678A.130G}, the zero-level intensities of our Effelsberg data need to be restored. 

While the restoration for the Cygnus X region has already been done in \citet{2023A&A...678A.130G}, the Galactic plane in the range $-2^{\circ}<\ell <60^{\circ}$ and $|b|<1^{\circ}$ remains uncorrected. For $10^{\circ}<\ell <60^{\circ}$, we continue to use the Urumqi 4.8~GHz continuum data \citep{2007A&A...463..993S, 2011A&A...527A..74S}. Since the Urumqi 4.8~GHz continuum data do not cover the area within $-2^{\circ}<\ell <10^{\circ}$, we instead used the Parkes 5~GHz data from the southern hemisphere survey of the Galactic plane \citep{1978AuJPA..45....1H} for the restoration in this region. Both the Urumqi and Parkes survey data are taken from the MPIfR's Survey Sampler\footnote{\url{https://www.mpifr-bonn.mpg.de/survey.html}}. These surveys cover overlapping regions, allowing us to compare their intensities. We convolved the data from both surveys to a common HPBW of 15\arcmin\, for comparison, as we only used these data for the large-scale restoration. 

We compare the two data sets within $10^{\circ}<\ell <30^{\circ}$ in Fig.~\ref{Fig:urmqi-parkes}. In Fig.~\ref{Fig:urmqi-parkes}a, we notice that the brightness temperatures of the Parkes survey data are  generally consistent with those of the Urumqi survey except for bright compact sources. Figure~\ref{Fig:urmqi-parkes}b suggests that most of the temperature differences lie between 0~K and 0.1~K with a median value of 0.028~K. We also observe that the temperature difference varies with Galactic latitude as is demonstrated for a Galactic longitude of $\ell =11\rlap{.}^{\circ}65$ (see Fig.~\ref{Fig:urmqi-parkes}c). The Galactic longitude at $\ell =11\rlap{.}^{\circ}65$ was selected for the purpose that it is predominantly characterized by extended emissions and largely devoid of bright compact emissions. To mitigate this large-scale variation, we applied a third-order polynomial fit to the Galactic-latitude-dependent variation. The fitting result was subsequently applied to the Parkes 5~GHz data within $-2^{\circ}<\ell<10^{\circ}$, which effectively matches the intensity scale between the two data sets. Finally, we combined the Urumqi data and the corrected Parkes data to provide a template of the large-scale distribution (see Fig.~\ref{Fig:template} in Appendix~\ref{app.template}). 

Following our previous studies \citep{2021A&A...651A..85B, 2023A&A...678A.130G}, we restored the zero-level distribution of the GLOSTAR 4.89~GHz continuum data using the large-scale template derived from the Urumqi and Parkes surveys. The restored Effelsberg 4.89~GHz continuum data were convolved to the Urumqi survey's HPBW (9\rlap{.}\arcmin5), and a comparison was made, as shown in Fig.~\ref{Fig:urumqi-glostar}. A linear fit to the observed data points reveals that the brightness temperature of the Urumqi survey is $\sim$97.4\% of that of the GLOSTAR survey, indicating a good agreement between the two surveys within 4\%. 

For the polarization data, corrections for zero-level offsets and baseline distortions are required. Unlike Stokes $I$, Stokes $Q$ and $U$ maps can contain negative values, rendering the background filtering method \citep{1979A&AS...38..251S} used for Stokes $I$ maps unsuitable. Instead, we applied a modified version of this method \citep{2002A&A...390..337K}, which accounts for positive and negative intensity values. For $\ell >10^{\circ}$, zero-level offsets were restored using Urumqi polarization data, while for $\ell <10^{\circ}$, we utilized the nine-year WMAP K-band data \citep{2013ApJS..208...20B}. We separated small- and large-scale components at 90\arcmin\,angular resolution in both the GLOSTAR and the WMAP maps. The large-scale WMAP component is scaled assuming a brightness temperature spectral index $\beta=-2.9$ \citep{1988A&AS...74....7R} and added to the GLOSTAR maps. Figure~\ref{Fig:pol-restore} presents the zero-restoration of the GLOSTAR 4.89~GHz polarization data in the Galactic center. The plot indicates that the large-scale Stokes $U$ and $Q$ distributions have been successfully corrected, while the small-scale structures remain nearly identical to the original ones. Hence, this approach effectively corrects zero-level offsets and mitigates baseline distortions in the Effelsberg polarization maps.

\section{Results and discussion}\label{Sec:res}
\subsection{Overall distribution}\label{sec.mor}

\begin{figure*}[!htbp]
\centering
\includegraphics[width = 0.95 \textwidth]{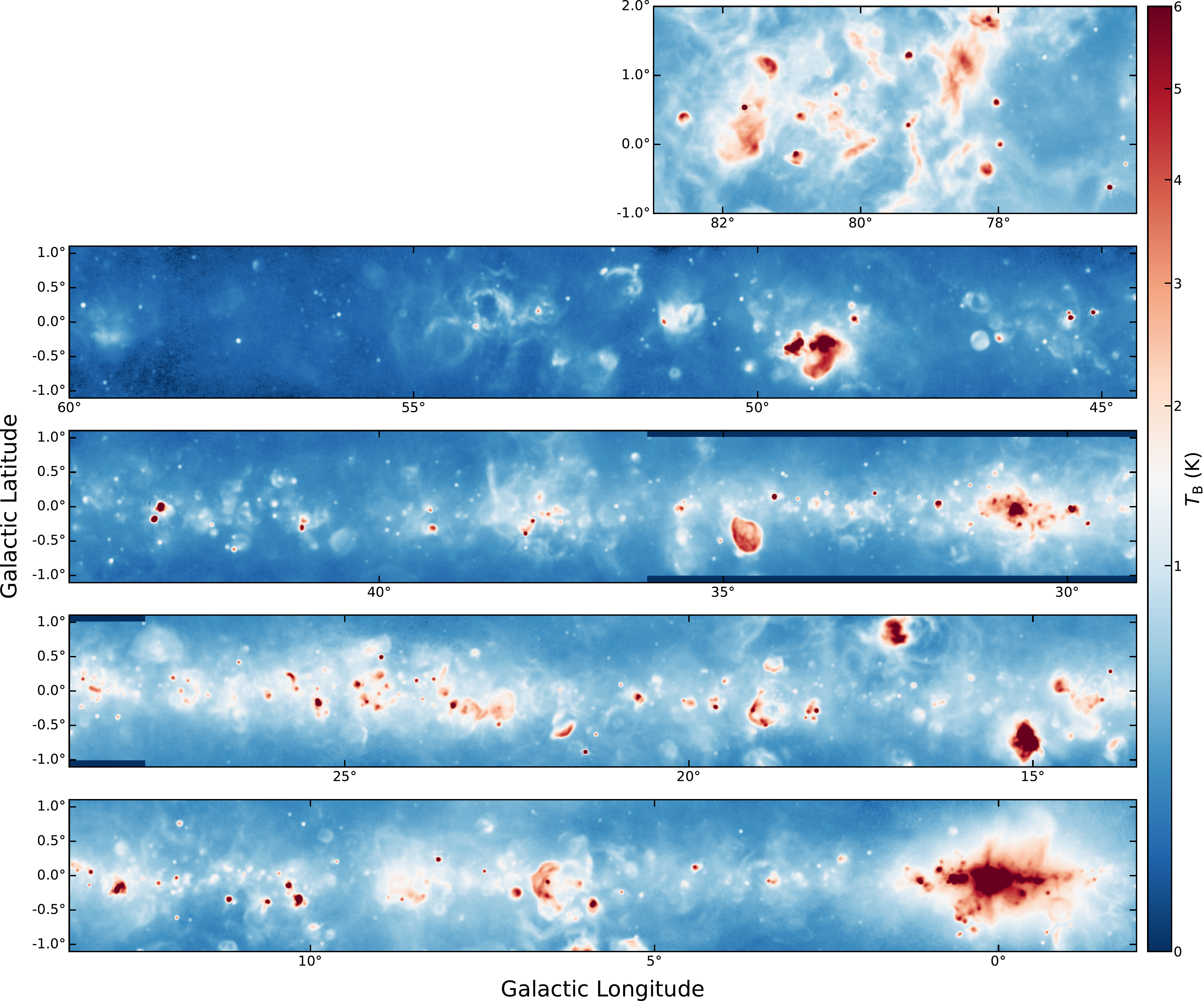}
\caption{{Effelsberg 4.89~GHz Stokes $I$ continuum map of the whole GLOSTAR survey. Zoom-in plots of each fields are available via \url{https://gongyan2444.github.io/glostar-snr-hii.html}, where the green, gray, and red circles represent the known SNRs \citep{2025JApA...46...14G}, SNR candidates \citep{2017A&A...605A..58A,2021A&A...651A..86D,2025A&A...693A.247A}, and H{\scriptsize II} regions from the WISE catalog \citep{2014ApJS..212....1A}, respectively.}\label{Fig:4.89GHz-I}}
\end{figure*}

\begin{figure*}[!htbp]
\centering
\includegraphics[width = 0.95 \textwidth]{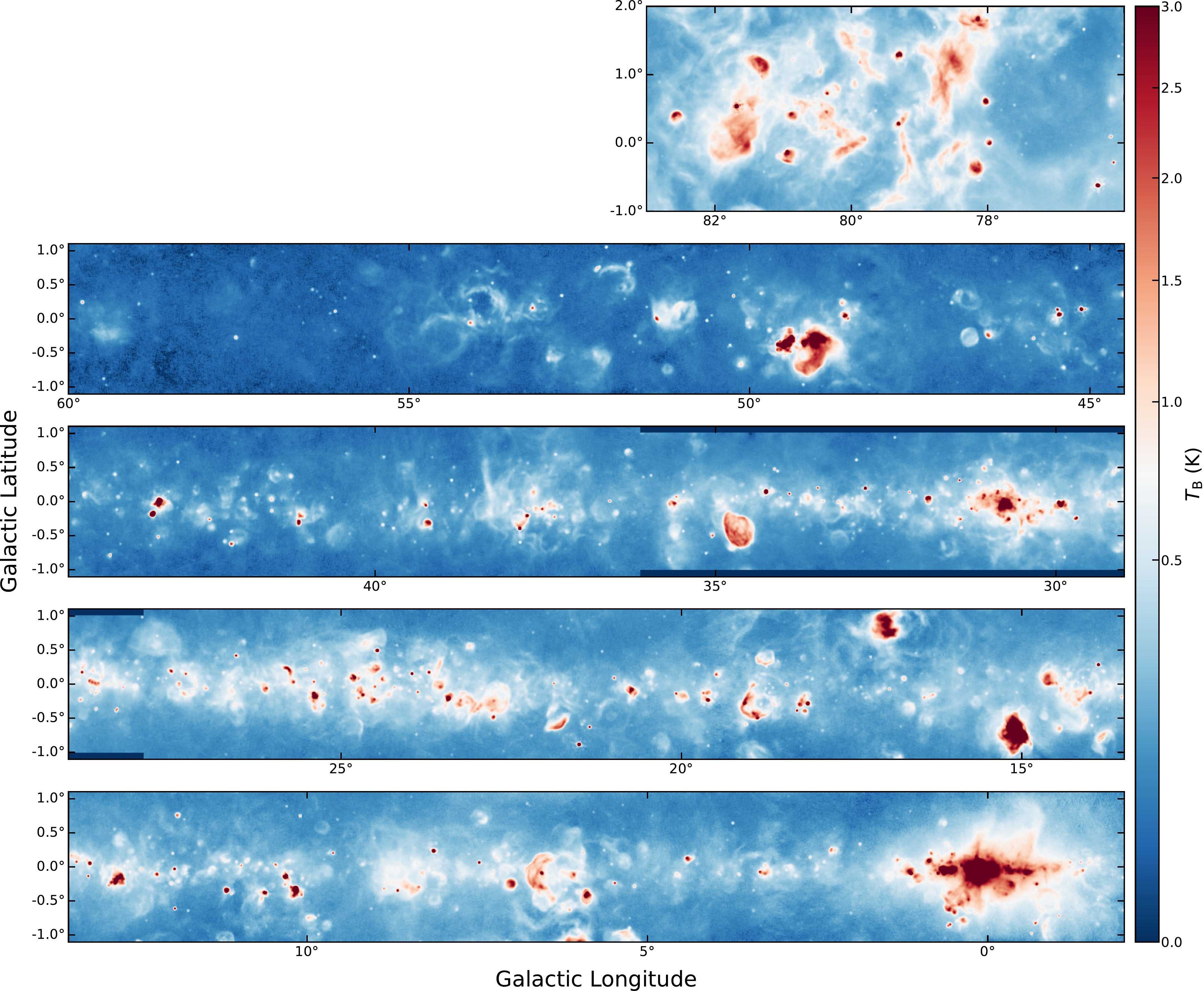}
\caption{{Effelsberg 6.82~GHz Stokes $I$ continuum map of the whole GLOSTAR survey.}\label{Fig:6.82GHz-I}}
\end{figure*}

Figures~\ref{Fig:4.89GHz-I} and \ref{Fig:6.82GHz-I} show the Effelsberg 4.89~GHz and 6.82~GHz Stokes $I$ data from the GLOSTAR survey. The overall distributions of Stokes $I$ continuum emissions closely resemble those observed in early radio continuum images \citep[e.g.,][]{Altenhoff70,1978AuJPA..45....1H,1990A&AS...85..633R,1990A&AS...83..539R,2007A&A...463..993S, 2011A&A...527A..74S}, but with higher angular resolution. The bulk of the bright emission is confined near the Galactic mid-plane and declines rapidly with increasing latitude (see the scale-height analysis in Appendix~\ref{app.height}), indicating that it predominantly traces structures within the Galactic disk. Numerous extended structures are evident throughout the maps. The brightest among them correspond to well-known H{\scriptsize II} regions and SNRs, which contribute to the thermal and non-thermal emission budgets, respectively (see Sect.~\ref{sec.extended} for a detailed discussion). These sources also play a role in shaping the observed asymmetry of the latitude profile (see Appendix~\ref{app.height}). Beyond these classical sources, the Effelsberg data also capture a variety of large-scale structures. These include, for example, the prominent bipolar radio chimney in the Galactic center, which extends to $\sim$430~pc above the mid-plane \citep[e.g.][]{2019Natur.573..235H,2023A&A...674L..15V}, as well as the bright and extended radio continuum emission associated with the Galaxy’s major mini-starburst regions, such as Sgr~B2, W43, and W51. In addition to these prominent objects, the Effelsberg data also reveal a wealth of compact extragalactic point sources, filaments, and diffuse features that contribute to the complex radio continuum landscape of the Galactic plane.

\begin{figure*}[!htbp]
\centering
\includegraphics[width = 0.95 \textwidth]{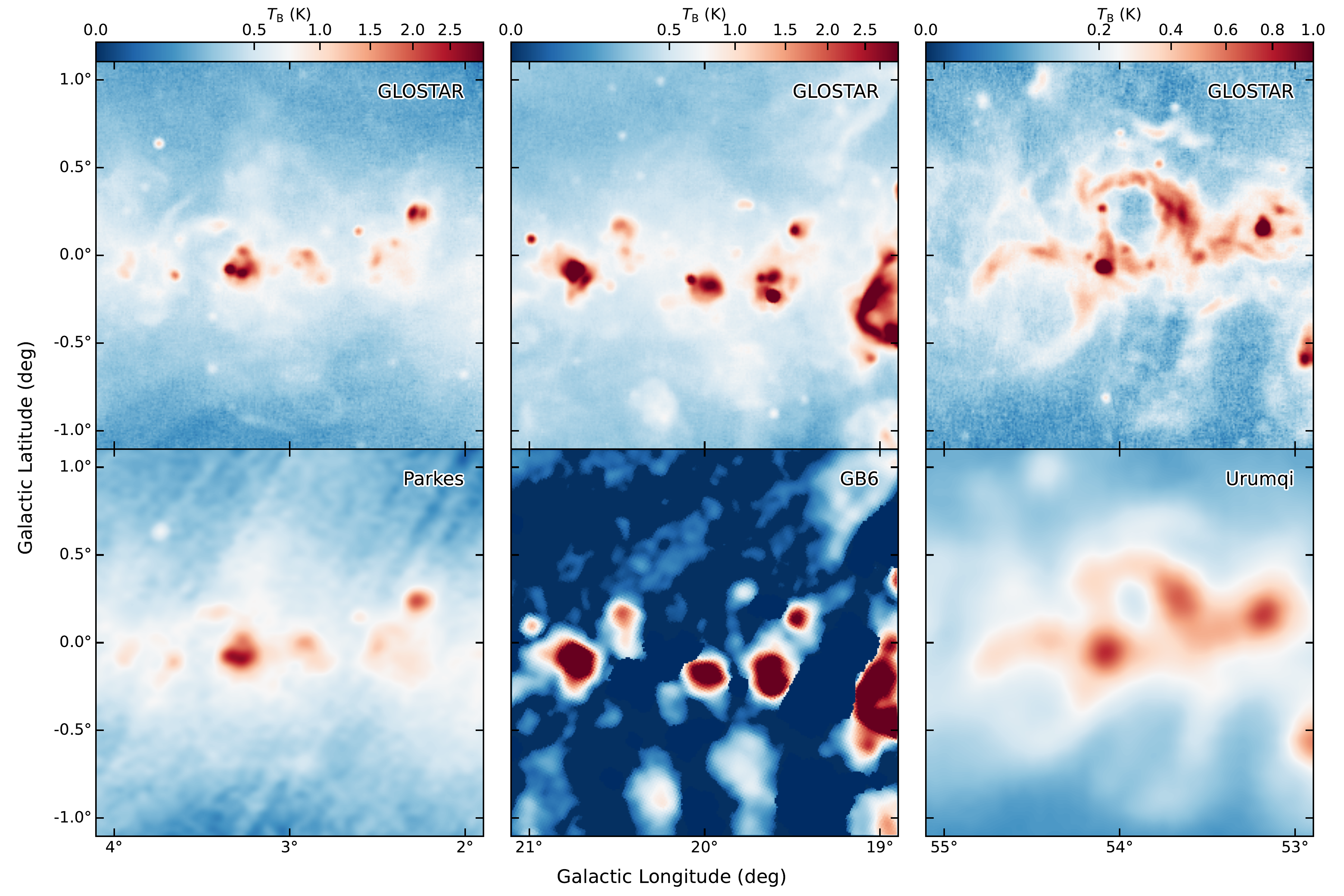}
\caption{{Comparison of the Effelsberg 4.89~GHz Stokes $I$ continuum map with those from the Urumqi, GB6, and Parkes surveys at similar frequencies.}\label{Fig:compare}}
\end{figure*}

To highlight the improvement in image fidelity of the GLOSTAR data compared to previous surveys at similar frequencies \citep{1978AuJPA..45....1H,1989AJ.....97.1064C,2007A&A...463..993S}, we show the Stokes $I$ continuum maps from different surveys in Fig.~\ref{Fig:compare}. Owing to the higher angular resolution and sensitivity, our GLOSTAR data reveal significantly finer structures and a larger number of faint sources compared to the Urumqi \citep{2007A&A...463..993S}, Green Bank 6~cm\footnote{From \url{https://skyview.gsfc.nasa.gov/current/cgi/query.pl}.} \citep[GB6;][]{1989AJ.....97.1064C}, and Parkes \citep{1978AuJPA..45....1H} radio continuum surveys at $\sim$5~GHz. Furthermore, our data exhibit fewer image artifacts than the Parkes continuum map, where scanning effects are evident (see the lower left panel of Fig.~\ref{Fig:compare}). The GB6 survey data suffer from zero-level offsets and uncorrected large-scale distortions, resulting in extensive regions of negative values in the survey map (see the lower middle panel of Fig.~\ref{Fig:compare}). These issues arise from the baseline subtraction method adopted in the GB6 survey \citep{1989AJ.....97.1064C}, which tends to suppress extended emission. This comparison demonstrates that the GLOSTAR radio continuum data are currently the highest-quality single-dish continuum data in this frequency range. 


\subsection{Missing flux: the importance of the Effelsberg 100-m observations}
One of the main goals of the Effelsberg radio continuum survey is to recover the missing flux in the radio interferometry data of the GLOSTAR survey, which lack the zero-spacing information. Given that our VLA observations have shortest baselines of $\sim$30~m (see Fig.~\ref{Fig:uv} for instance), our Effelsberg 100-m observations can perfectly fill the missing zero-spacing gap in the \textit{uv}-plane.

\begin{figure*}[!htbp]
\centering
\includegraphics[width = 0.95 \textwidth]{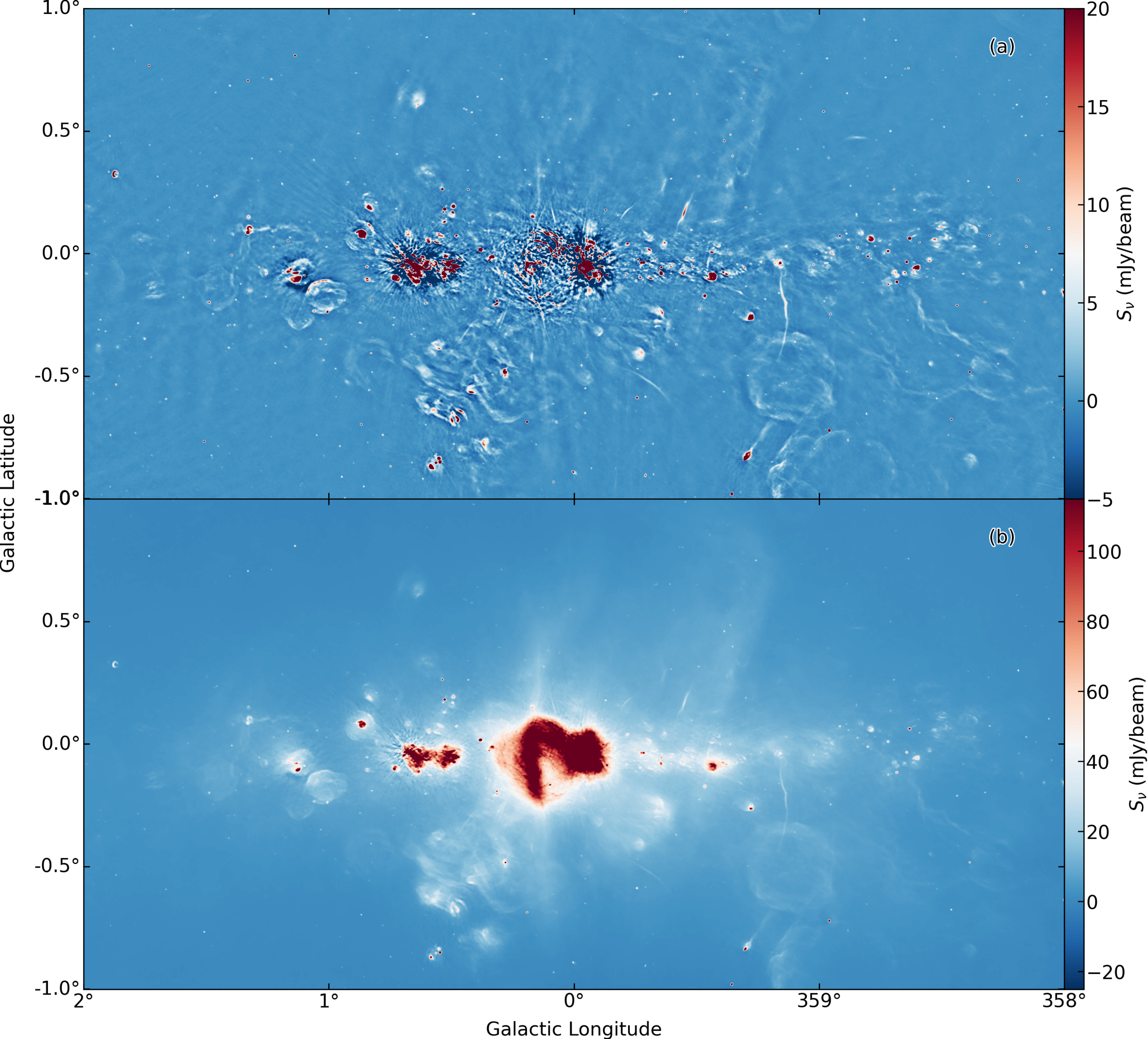}
\caption{{Comparison of the GLOSTAR data with and without incorporating the Effelsberg data at 5.8~GHz. (a): VLA D-array only continuum image. (b): Combination of the VLA D-array and the Effelsberg 100-m single-dish image.}\label{Fig:mf}}
\end{figure*}

We use the Stokes $I$ continuum distribution of the Galactic center to illustrate the improvement incorporating the Effelsberg 100-m data. As shown in Fig.~\ref{Fig:mf}a (see also Fig.~1 in \citet{2021A&A...651A..88N}), our VLA D configuration data have a maximum recoverable scale of $\sim$4\arcmin, resulting in a fragmented image. This fragmentation arises because the data primarily capture compact features, while the broader, diffuse emission is not adequately sampled in the VLA D configuration observations. To mitigate this issue, we combine the Effelsberg 100-m data with the VLA D configuration data following the method as described in \citet{2023A&A...671A.145D}. The resulting image, displayed in Fig.\ref{Fig:mf}b, presents a smooth, continuous distribution, revealing the full extent of the structure across all angular scales above $\sim$18\arcsec, which is the angular resolution of the combined data. Consequently, the brightness distribution more accurately traces extended sources in the Galactic plane.

\begin{figure*}[!htbp]
\centering
\includegraphics[width = 0.95 \textwidth]{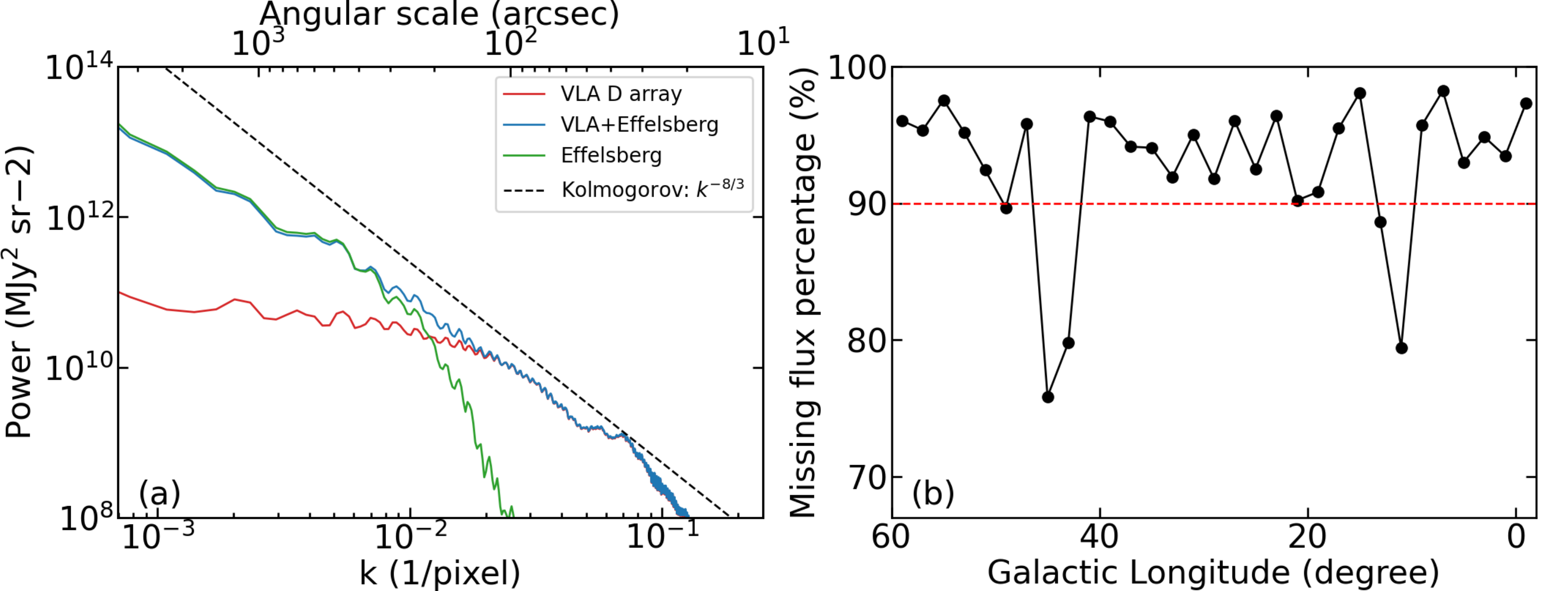}
\caption{{(a): Angular power spectra of the VLA-D array, Effelsberg-100~m, and VLA-D array+Effelsberg-100~m combined images toward the representative region near the Galactic center (see Fig.~\ref{Fig:mf}). The 2D Kolmogorov-like spectrum is indicated by the black dashed line for comparison. (b): Percentage of missing flux of the VLA D-array data across the Galactic plane covered by the GLOSTAR survey. The horizontal dashed red line indicates a missing flux fraction of 10\%.}\label{Fig:mf2}}
\end{figure*}

We construct the angular power spectra for both data sets to examine the distribution of power across different angular scales. The resulting spectra toward the Galactic center are presented in Fig.~\ref{Fig:mf2}a. The two power spectra agree well on angular scales  $\lesssim 2$\arcmin. On larger scales, however, the VLA-only image shows a slight deficit of power relative to the VLA+Effelsberg image. This indicates that, despite the nominal $\sim$4\arcmin\,largest angular scale of the D-array, the practical sensitivity to diffuse emission is reduced because of sparse short-baseline coverage and the snapshot uv sampling. Compared with the VLA-only image, the spatial power spectrum of the VLA+Effelsberg image is better described with a single power-law distribution on angular scales of $\gtrsim$1\arcmin, showing that the power of the ionized gas decreases toward smaller angular scales. A linear fit to the data for angular scales $>$100\arcsec\,yields a power index of about $-$2.3, which is slightly shallower than the 2D Kolmogorov-like spectral index of $-8/3$ predicted for isotropic, incompressible turbulence \citep{1941DoSSR..30..301K}. Previous studies indicate that electron density fluctuations associated with interstellar turbulence are likely to follow a Kolmogorov-like spectrum \citep{1995ApJ...443..209A}, which describes an energy cascade from large to small scales. The slightly flatter spectrum observed here may indicate the non-negligible contribution of small-scale turbulence to the overall spectrum \citep{2000ApJ...537..720L}.

Using the integrated flux densities over each field (see Sect.~\ref{Sec:obs}) in both VLA-only and VLA+Effelsberg images, we estimate the total flux density and thereby quantify the percentage of flux density missed by the VLA data. Comparing the derived values in Fig.~\ref{Fig:mf2}b, we find that nearly all fields miss more than 90\% of flux densities in the VLA D array images. This highlights the crucial role of the Effelsberg 100-m observations in recovering the extended emission for the GLOSTAR survey.




\subsection{Polarization}
\begin{figure*}[!htbp]
\centering
\includegraphics[width = 1.0 \textwidth]{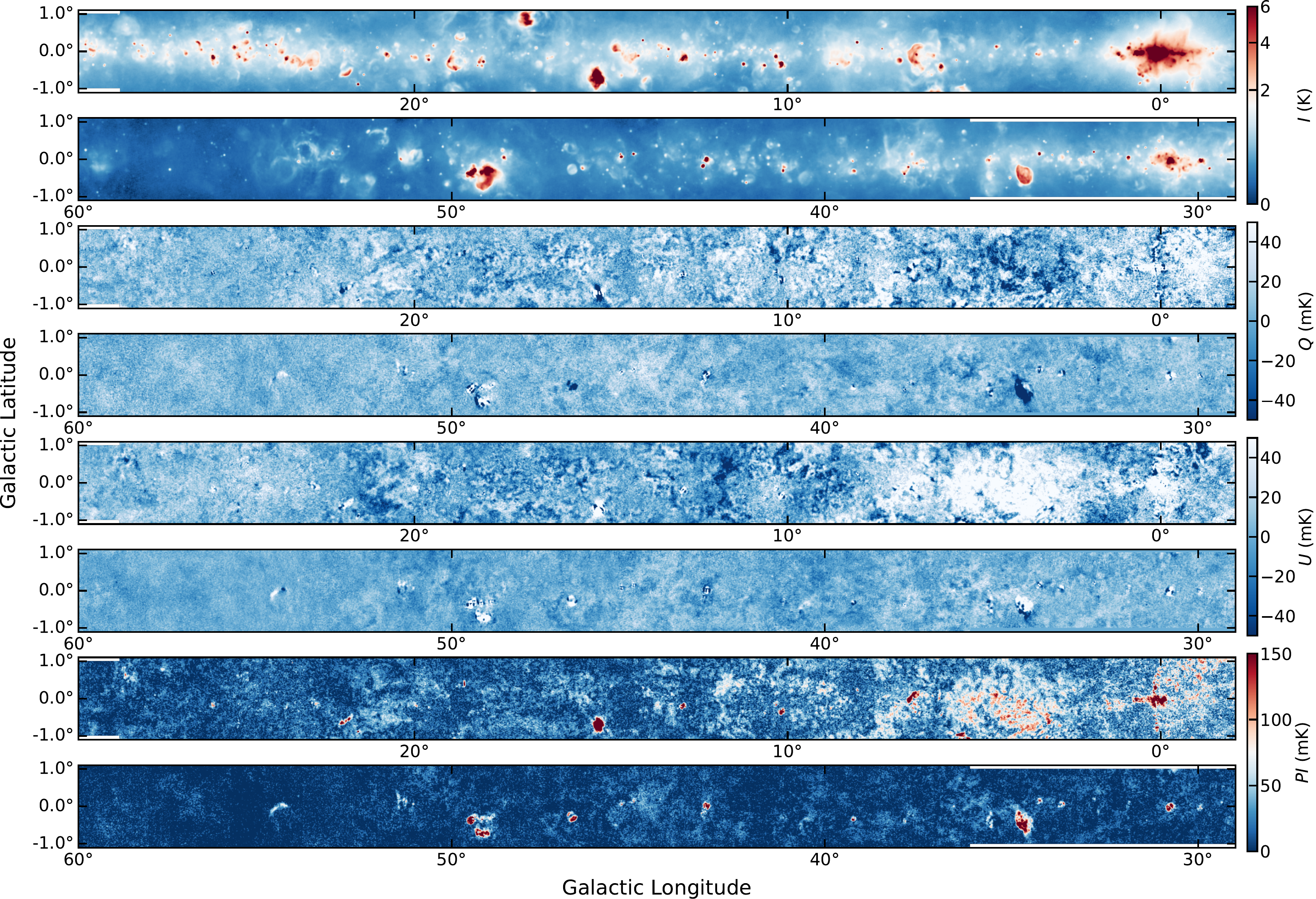}
\caption{{Distribution of the Stokes $I$, $Q$, $U$, and polarization intensity maps from the Effelsberg 4.89~GHz continuum emission of the GLOSTAR survey. Zoom-in plots of each fields are available via \url{https://gongyan2444.github.io/glostar-snr-hii.html}, where the green, gray, and red circles represent the known SNRs \citep{2025JApA...46...14G}, SNR candidates \citep{2017A&A...605A..58A,2021A&A...651A..86D,2025A&A...693A.247A}, and H{\scriptsize II} regions from the WISE catalog \citep{2014ApJS..212....1A}, respectively.}\label{Fig:4.89GHz-pol-all}}
\end{figure*}

\begin{figure*}[!htbp]
\centering
\includegraphics[width = 1.0 \textwidth]{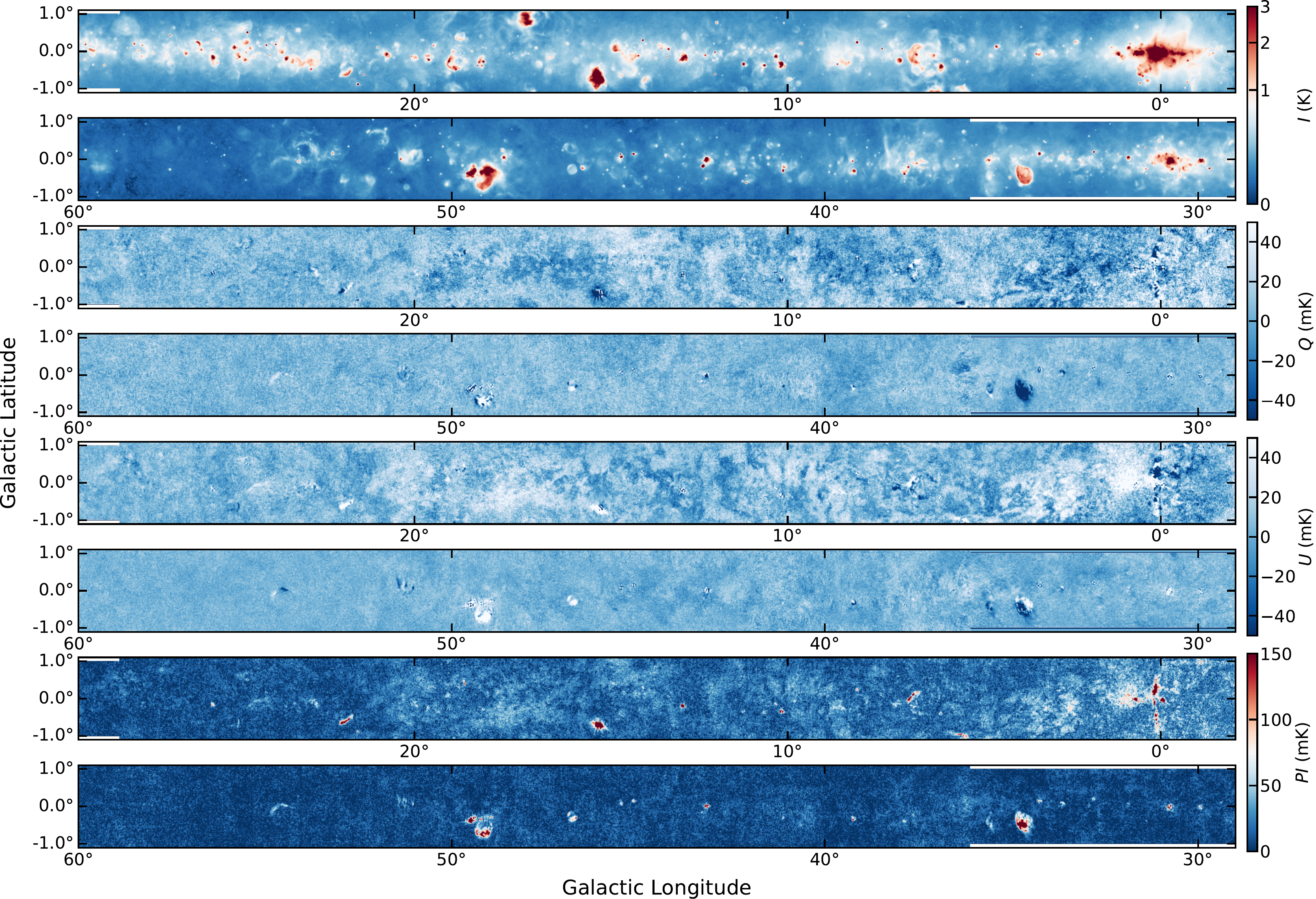}
\caption{Distribution of the Stokes $I$, $Q$, $U$, and polarization intensity maps from the Effelsberg 6.82~GHz continuum emission of the GLOSTAR survey.}\label{Fig:6.82GHz-pol-all}
\end{figure*}

Polarization data from radio continuum observations provide valuable insights into magnetic field structures in astrophysical objects. Figures~\ref{Fig:4.89GHz-pol-all} and \ref{Fig:6.82GHz-pol-all} present the Effelsberg 4.89~GHz and 6.82~GHz full Stokes data from the GLOSTAR survey. Compared to previous polarized continuum surveys at similar frequencies \citep{2007A&A...463..993S,2019MNRAS.490.2958J}, our Effelsberg polarization data provide an improvement in angular resolution by at least a factor of three. 

Consistent with previous polarization surveys \citep[e.g.][]{2007A&A...463..993S}, our observations reveal widespread diffuse polarized structures that often lack corresponding counterparts in Stokes~$I$. This effect is particularly pronounced toward the inner Galaxy, where the polarized emission becomes highly patchy. Two main classes of mechanisms have been proposed to explain such behavior \citep[e.g.][]{2014MNRAS.437.2936S}: (i) intrinsically irregular emission produced in regions dominated by random magnetic fields \citep[e.g.][]{2011A&A...527A..74S}, or (ii) the modulation of a smooth background by foreground Faraday screens. The latter may arise either from MHD turbulence within the warm ionized medium \citep{2011Natur.478..214G} or from discrete ionized structures such as H{\scriptsize II} regions \citep{2001ApJ...549..959G,2011A&A...527A..74S}. In our data (e.g. near $\ell \sim 9^{\circ}$; see Figs.~\ref{Fig:4.89GHz-pol-all} and \ref{Fig:6.82GHz-pol-all}), the polarized emission appears more fragmented at 4.89~GHz than at 6.82~GHz, consistent with stronger depolarization at the lower frequency and thus supporting the foreground-screen scenario. Nevertheless, dedicated studies will be required to robustly quantify the underlying mechanisms and to constrain the physical origin of these depolarizing structures.

From our data, we detect prominent polarization structures in at least 30 SNRs and SNR candidates. For five sources (e.g., G8.8583$-$0.2583, G21.0$-$0.4, G22.7$-$0.2, G26.75+0.73, and G54.4$-$0.3) that previously showed polarization only in VLA D-array data \citep{2021A&A...651A..86D}, our Effelsberg observations deliver the first single-dish detections of polarization. Although the VLA data reveal small-scale polarized structures, they are insensitive to the extended polarized emission that is filtered out by the interferometer. Our Effelsberg measurements therefore play a crucial role in recovering the large-scale polarization. Even for SNRs with earlier polarization measurements at lower frequencies, our observations offer improved constraints on the intrinsic polarization structures (see Sect.~\ref{sec.snr} for instance). In addition, our observations also reveal extended polarization structures that lack corresponding features in Stokes~$I$, in agreement with previous polarization surveys \citep[e.g.,][]{2007A&A...463..993S}. Such ``polarization-only" features are commonly attributed to the diffuse Galactic synchrotron emission, with their morphology and visibility shaped by Faraday rotation and depolarization effects along the line of sight.




%

It should be noted that the instrumental polarization has not been completely removed from the polarization images (see Sect.~\ref{Sec:eff} for details on the instrumental polarization measurements). For example, a source with a brightness temperature of $T_\mathrm{b} = 1$~K can contribute up to 10~mK to the polarized intensity maps, assuming an instrumental polarization level of 1\%. Consequently, leakage from bright H{\scriptsize II} regions may introduce a bias in the polarization measurements. By cross-checking the Stokes $I$ and polarization images, we identify a few H{\scriptsize II} regions where such contamination may significantly affect the polarization results. These include G005.887$-$00.443, G008.137+00.232, W31, G010.308$-$00.150, W33, G013.880+00.285, M16, G018.144$-$00.281, G019.609$-$00.239, M17, G020.728$-$00.105, G023.957+00.149, G025.382$-$00.151, G29, W43, G032.800+00.190, NRAO584, G037.763$-$00.212, W47, W49, G045.121+00.133, K47, W51\footnote{Although the polarization of W51C (also known as G49.2$-$0.7) is located in the southern part of W51, this SNR shows clear polarization. Its polarization properties have previously been studied at both low angular resolutions \citep{1974A&A....32..375V} and high angular resolution with VLA D-array data \citep{2021A&A...651A..86D}. Given the extended nature of the polarized emission, our Effelsberg observations are crucial for recovering the flux missed by the VLA data.}, S106, DR7, DR15, DR22, and DR21. We therefore caution that the polarization analysis in these regions should be interpreted with care. 


\subsection{Extended objects}\label{sec.extended}
As shown in the online version of Fig.~\ref{Fig:4.89GHz-I}, the GLOSTAR data encompass 99 known SNRs \citep{2025JApA...46...14G}, 241 SNR candidates \citep{2017A&A...605A..58A,2021A&A...651A..86D,2025A&A...693A.247A}, and 3633 H{\scriptsize II} regions from the WISE catalog \citep{2014ApJS..212....1A}, suggesting that discrete radio continuum sources are dominated by SNRs and H{\scriptsize II} regions in the Galactic plane. H{\scriptsize II} regions emit thermal free-free radiation from ionized gas surrounding young massive stars. In contrast, SNRs are typically characterized by non-thermal synchrotron emission. The morphology and spectral index derived from radio continuum observations provide key diagnostics of the physical conditions and evolutionary stages of both H{\scriptsize II} regions and SNRs, while the magnetic field structure inferred from polarization measurements offers additional insights into the latter. The GLOSTAR Effelsberg survey thus provides a valuable dataset for detailed investigations of the nature and properties of these populations. In the following, we present case studies of representative examples from each category to demonstrate the excellent capability of the Effelsberg data.

\subsubsection{H{\scriptsize II} regions}
We selected G016.648$-$00.357 (also known as Sh 2-48) as a representative case to demonstrate our analysis methodology toward an H{\scriptsize II} region. At a distance of 3.8~kpc \citep[e.g.,][]{2013A&A...556A.105O,2021PASJ...73S.368T}, this evolved H{\scriptsize II} region is extended with a radius of $\sim$7\arcmin\,(i.e., 7.7~pc), making our Effelsberg data more suitable for analysis compared to the GLOSTAR VLA observations. 

\begin{figure*}[!htbp]
\centering
\includegraphics[width = 0.95 \textwidth]{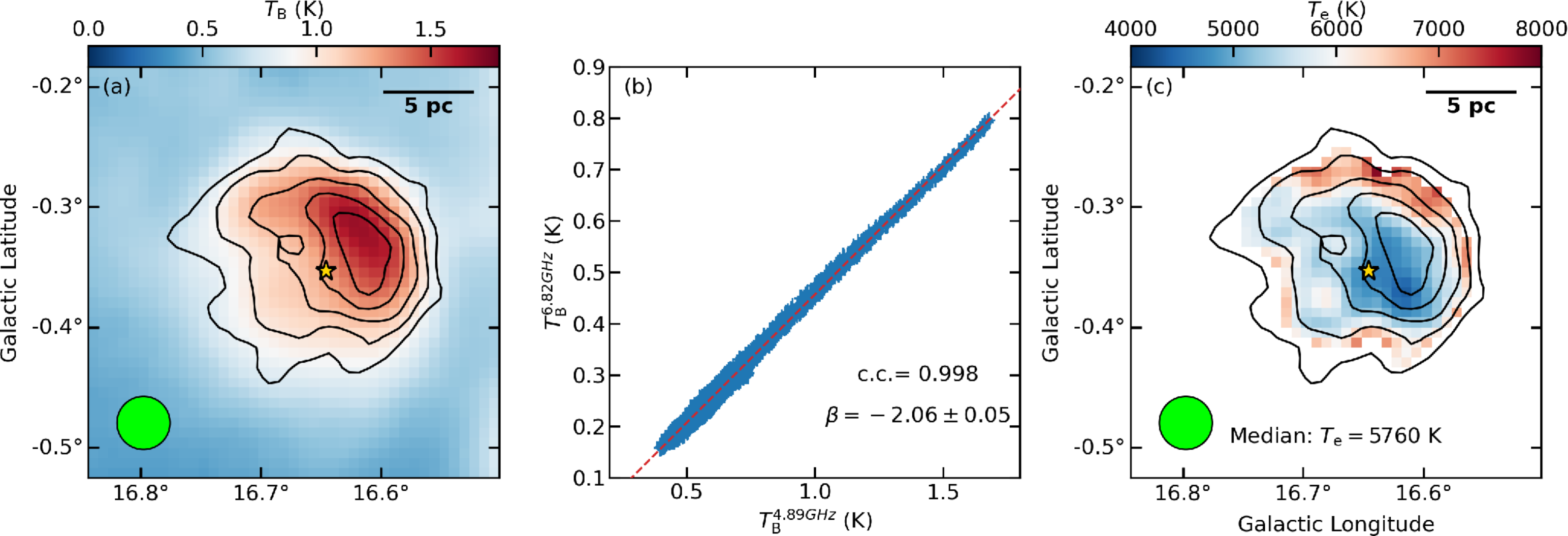}
\caption{{Observed properties of G016.648$-$00.357 (i.e., Sh 2-48). (a) Effelsberg 4.89~GHz Stokes $I$ image overlaid with the GDIGS RRL integrated intensity contours. (b) Temperature-Temperature plot between 4.89~GHz and 6.82~GHz at a common angular resolution of 2\rlap{.}\arcmin65. The red dashed line shows the best fit to the observed data points. (c) Spatial distribution of the derived electron temperatures overlaid with the GDIGS RRL integrated intensity contours. In panels (a) and (c), the integrated velocity range is from 0 to 70 \kms\,for the GDIGS RRL integrated intensity contours, starting at 1~K~\kms\,and increase by 0.5~K~\kms. The location of the exciting star BD$-$14~5014 is marked with a gold star. In each panel, the beam is shown in the lower-left corner, and the scale bar in the top-right corner is based on an assumed distance of 3.8~kpc \citep[e.g.,][]{2013A&A...556A.105O,2021PASJ...73S.368T}.}\label{Fig:S48}}
\end{figure*}

Figure~\ref{Fig:S48}a presents the radio continuum distribution of this region, which exhibits a cometary morphology, with a bright head located toward the northwest and a diffuse tail extending toward the southeast. No polarized emission is detected in this region. To constrain the spectral index\footnote{The spectral index $\alpha$, defined as $S_{\nu}\sim \nu^{\alpha}$, can be related to the Rayleigh–Jeans approximation, where $S_{\nu}\sim \nu^{2} T_{\rm B}$. As the brightness temperature spectral index $\beta$ is defined as $T_{\rm B}\sim \nu^{\beta}$, this yields $\alpha=\beta+2$.}, we employed a temperature–temperature (TT) plot, which is not affected by zero-level uncertainties in the images \citep[e.g.,][]{1962MNRAS.124..297T}. The TT plot, shown in Fig.\ref{Fig:S48}b, reveals a linear relation between the radio continuum emission at 4.89~GHz and 6.82~GHz, with a high Pearson correlation coefficient of 0.997. A linear fit to the observed data yields a brightness temperature spectral index of $\beta = -2.06 \pm 0.05$, consistent with an optically thin spectral index for H{\scriptsize II} regions \citep[e.g.,][]{2019A&A...623A.105G}. This result indicates that continuum emissions at both bands are in excellent agreement with  optically thin conditions.

Electron temperatures are a fundamental parameter for characterizing the physical conditions in H{\scriptsize II} regions. Its accurate determination requires measurements of both the radio continuum and associated radio recombination line (RRL) observations. In particular, our simultaneous RRL and continuum observations enables investigations into spatial fluctuations of the electron temperature within H{\scriptsize II} regions. To derive electron temperatures, we combined our continuum maps with RRL data from the GBT Diffuse Ionized Gas Survey \citep[GDIGS\footnote{\url{http://astro.phys.wvu.edu/gdigs/}};][]{2021ApJS..254...28A}, which provides Hn$\alpha$ measurements at a mean frequency of 5.69~GHz and an angular resolution of 2\rlap{.}\arcmin65. To match this resolution, we convolved our radio continuum data at both frequencies to 2\rlap{.}\arcmin65. The continuum emission at 5.69~GHz was then derived by interpolating our convolved data using the brightness temperature spectral index $\beta = -2.06$. Assuming optically thin emission and adopting a main beam efficiency of 92\%\footnote{\url{https://www.gb.nrao.edu/GBT/DA/gbtidl/gbtidl_calibration.pdf}} for the GDIGS data, electron temperatures, $T_{\rm e}$, were calculated following the relation \citep[e.g.,][]{2019ApJ...887..114W}:
\begin{equation}\label{f.te}
\left(\frac{T_{\rm e}}{\rm K}\right)  = \left\{7100 \left(\frac{\nu}{\rm GHz}\right)^{1.1}
\left(\frac{S_{\rm C}}{S_{\rm L}}\right)
\left(\frac{\Delta \varv}{\rm km\; s^{-1}}\right)^{-1} 
 \left[1+\frac{n(^4{\rm He}^{+})}{n({\rm H}^{+})}\right]^{-1} \right\}^{0.87} 
\end{equation}
where $\frac{S_{\rm C}}{S_{\rm L}}$ is the continuum-to-line ratio, $\Delta \varv$ is the FWHM line width, and $n(^4{\rm He}^{+})/n({\rm H}^{+})$ is the ${\rm He}^{+}/{\rm H}^{+}$ abundance ratio which was assumed to be 0.08 \citep[e.g.,][]{2006ApJ...653.1226Q,2015A&A...581A..48G}. 

The resulting distribution of electron temperatures is shown in Fig.~\ref{Fig:S48}c, spanning 4351~K to 7819~K, with a median of 5685~K. The associated 1$\sigma$ uncertainties range from 193~K to 813~K, with a median uncertainty of 418~K. These values are slightly lower than those predicted by the Galactic electron temperature gradient \citep[e.g.,][]{2024A&A...689A..81K}.
In this figure, elevated electron temperatures are preferentially found in the boundary of the H{\scriptsize II} region, exhibiting a gradient that decreases towards the center. Such a temperature structure is likely driven by photon hardening effects at the edges (see Figure~7.9 in \citealt{2005pcim.book.....T} for instance; \citealt{2022A&A...664A.140K}), where lower-energy photons are preferentially absorbed by neutral hydrogen due to their larger ionization cross sections, allowing higher-energy photons to penetrate further into ambient gas.



\subsubsection{Supernova remnant}\label{sec.snr}
\begin{figure*}[!htbp]
\centering
\includegraphics[width = 1.0 \textwidth]{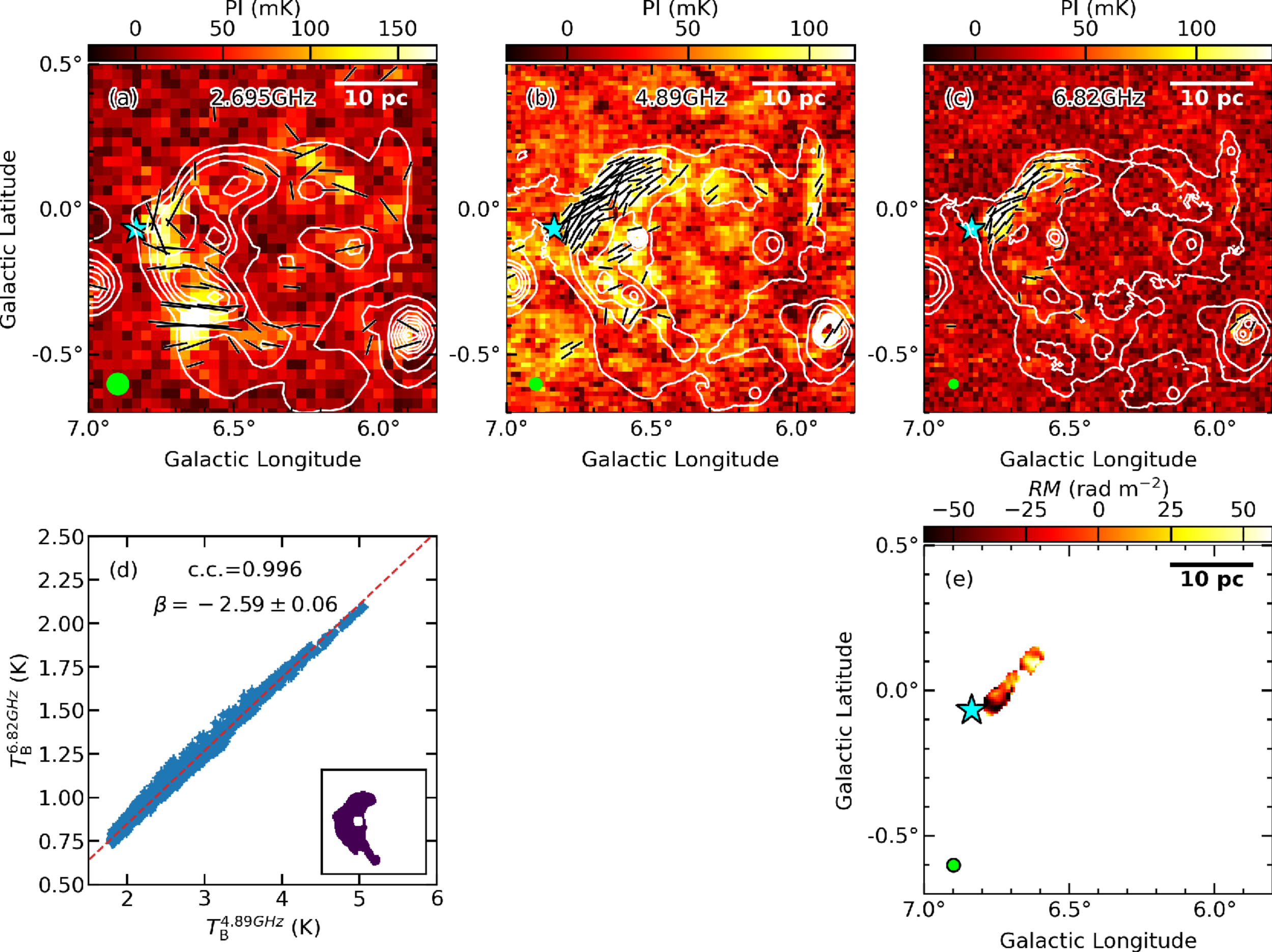}
\caption{{GLOSTAR radio continuum data of W28. (a) Linear polarization intensity maps overlaid with the magnetic field vectors, derived from the 2.695~GHz Effelsberg Galactic plane survey. The white contours represent the 2.695~GHz Stokes $I$ continuum emission, starting at 2.5~K and increasing in steps of 2.5~K. (b) Similar to panel (a), but using the GLOSTAR 4.89 GHz continuum data, with white contours starting at 1.0~K and increasing in steps of 1.0~K. (c) Similar to panel (a), but based on the GLOSTAR 6.82 GHz continuum data, where white contours start at 0.5 K and increase in steps of 0.5 K. In panels (a)--(c), the polarization angles have been rotated by 90$^{\circ}$ to trace the magnetic field directions which are indicated by black bars. The position of the runaway pulsar PSR~B1758$-$23 (i.e., PSR~J1801$-$23), thought to be associated with W28 \citep{1993Natur.365..136F}, is indicated by a cyan star. In each panel, the beam is shown in the lower-left corner, and the scale bar in the top-right corner is based on an assumed distance of 2~kpc \citep{2002AJ....124.2145V}. (d) Temperature-Temperature plot of the northeastern shell between 4.89~GHz and 6.82~GHz at a common angular resolution of 160\arcsec. The red dashed line shows the best fit to the observed data points which are extracted from the region displayed in the panel in the lower right corner. (e) Distribution of the derived rotation measures.}\label{Fig:w28}} 
\end{figure*}

As an example for extended SNRs from our survey, we present maps of SNR W28 (also known as G6.4$-$0.1) that is a well-known mixed-morphology SNR associated with very high energy $\gamma$-ray sources \citep{2008A&A...481..401A} and a runaway pulsar \citep{1993Natur.365..136F}. In particular, polarization measurements of W28 are rare, with only a few early studies available (e.g., see Fig.~9b in \citealt{1972A&A....20..237K} and Fig.~15 in \citealt{1976AuJPh..29..435D}). In Fig.~\ref{Fig:w28}, we present a comparison of the polarization results toward W28 from the 2.695~GHz Effelsberg Galactic plane survey \citep{1999A&A...350..447D} and our GLOSTAR survey. Strong and ordered linear polarization is detected along its radio shell at 4.89~GHz and 6.82~GHz in our GLOSTAR data, with the inferred magnetic field vectors (obtained by rotating the observed polarization vectors by 90\degree) oriented nearly parallel to the shell and exhibiting a low dispersion in polarization position angles. This tangential distribution is consistent with the expected compression of magnetic fields by the supernova shock, a phenomenon commonly observed in evolved SNR shells \citep[e.g.,][]{1992A&A...256..214R,2004mim..proc..141F,2017ARA&A..55..111H}. Furthermore, the tangential pattern observed in our data is more pronounced than that seen in polarization measurements from the 2.695~GHz Effelsberg Galactic plane survey (see the left panel of Fig.~\ref{Fig:w28}) and previous studies \citep{1972A&A....20..237K,1976AuJPh..29..435D}, indicating significant depolarization effects at lower frequencies. 

We also employed the TT plot method to derive the spectral index of the source. W28 is known to be superimposed with several H{\scriptsize II} regions along the line of sight \citep[e.g.,][]{2014ApJS..212....1A}. To minimize contamination from thermal emission, we therefore restricted our analysis to the northeastern shell to determine the spectral index. As shown in Fig.~\ref{Fig:w28}d, a clear linear relation is observed between the two bands, with a high Pearson correlation coefficient of 0.996. A linear fit to the data yields a brightness temperature spectral index of $\beta = -2.59 \pm 0.06$ (i.e., $\alpha = -0.59 \pm 0.06$), which is much steeper than the value of $\alpha = -0.35 \pm 0.18$ reported at lower frequencies \citep{2000AJ....120.1933D}. This difference could be due to our choice of region, as the northeastern shell is less affected by emission from nearby H{\scriptsize II} regions compared to previous studies.

The interquartile range of the linear polarization degrees span from 6\% to 10\%\,with a peak value of 15\%\,at 4.89~GHz and from 6\% to 12\%\,with a peak value of 17\%\,at 6.82~GHz, with both frequencies exhibiting a median value of 8\%. These results indicate consistently high levels of linear polarization fractions across the two bands. According to the synchrotron theory \citep[e.g.,][]{2013tra..book.....W}, the intrinsic degree of linear polarization of synchrotron radiation emitted by relativistic electrons spiraling in a uniform magnetic field is given by
\begin{equation}
    \Pi_{0} = \frac{3-3\alpha}{5-3\alpha}\;,
\end{equation}
where $\alpha=\beta+2$. Adopting a brightness temperature spectral index of $\beta = -2.59$, we obtained $\Pi_{0}=70\%$, which is much higher than our measured values. If the reduction in the observed polarization degree is attributed to the presence of a random magnetic field component, the observed linear polarization degree can be expressed as 
\begin{equation}
    \Pi = \Pi_{0}\frac{H_{\rm u}^{2}}{H_{\rm u}^{2}+H_{\rm r}^{2}} \;,
\end{equation}
where $H_{\rm u}$ and $H_{\rm r}$ are the uniform and random field strengths \citep[e.g.,][]{1966MNRAS.133...67B}, respectively. Using a median value of 8\%\,for the observed degree of linear polarization, we can estimate the uniform-to-random field ratios, $\frac{H_{\rm u}^{2}}{H_{\rm r}^{2}}$, to be about 0.13. For a maximum $\Pi=15\%$, $\frac{H_{\rm u}^{2}}{H_{\rm r}^{2}}$ increases to 0.27. If Faraday depolarization also contributes to the reduced degree of linear polarization, the estimated $\frac{H_{\rm u}^{2}}{H_{\rm r}^{2}}$ values should be taken as lower limits. 


Our survey provides polarization data at two frequency bands, enabling the determination of rotation measures (RMs). We convolved the Stokes $Q$ and $U$ maps from both bands to a common angular resolution of 160\arcsec, and re-calculated the polarization angles. Neglecting the $n\pi$ ambiguity arising from the indistinguishable orientation of the vector under a 180\degree\,(i.e., $\pi$) rotation, we derived the RMs directly from the polarization angles at the two frequencies, and the RM map is shown in Fig.~\ref{Fig:w28}e. The resulting RMs range from $-61.0$ to $44.1$~rad~m$^{-2}$ in the northeastern shell of W28, which are significantly lower than those measured in the textbook mixed-morphology SNR W44 \citep[e.g.,][]{2011A&A...536A..83S,2017MNRAS.464.4107G}. This may indicate lower electron densities, weaker line-of-sight magnetic fields, or differences in the Galactic RM environments. 


As shown above, our results represent a significant improvement over earlier polarization studies, with the GLOSTAR observations providing substantially higher quality single-dish data in tracing the intrinsic polarization properties. This enhancement enables a more detailed and reliable investigation of the polarization and magnetic field structures within the Milky Way.


\section{Summary and conclusion}\label{Sec:sum}
As part of the GLOSTAR survey project, we present the large-scale radio continuum observations covering the Galactic plane in the range $-2^{\circ}<\ell <60^{\circ}$ and $|b|<1^{\circ}$, as well as a section in Cygnus X ($76^{\circ}<\ell <83^{\circ}$ and $-1^{\circ}<b<2^{\circ}$) with the Effelsberg 100-m radio telescope. The GLOSTAR dataset surpasses the quality of previous surveys at similar frequencies, including those from the Urumqi, Parkes, GB6, and early Effelsberg surveys. Furthermore, we show that the Effelsberg continuum data are crucial for tracing the large-scale structure of ionized gas in the Milky Way. Our polarization data, less affected by depolarization effects compared to those at lower frequencies, offer reliable tracers of the intrinsic magnetic field orientation. These radio continuum data are freely available to the scientific community via \url{https://glostar.mpifr-bonn.mpg.de/glostar/} and \url{https://www.mpifr-bonn.mpg.de/survey.html}. The data are provided in FITS and other formats, facilitating their use for the investigation of extended radio continuum emission in the Milky Way. The VLA+Effelsberg combined data which are valuable for studying extended structures with scales $\gtrsim$18\arcsec\,will be released in a forthcoming publication. These valuable dataset will significantly contribute to future studies of astrophysical objects. 

\section*{ACKNOWLEDGMENTS}\label{sec.ack}
We thank the Effelsberg 100-m telescope staff for their assistance with our observations. YG was supported by the Ministry of Science and Technology of China under the National Key R\&D Program (Grant No. 2023YFA1608200), the National Natural Science Foundation of China (Grant No. 12427901), and the Strategic Priority Research Program of the Chinese Academy of Sciences (Grant No. XDB0800301). AYY acknowledges the support from the National Key R$\&$D Program of China No.~2023YFC2206403 and National Natural Science Foundation of China (NSFC) grants No. 12303031 and No. 11988101. S.A.D. acknowledges the M2FINDERS project from the European Research Council (ERC) under the European Union's Horizon 2020 research and innovation programme (grant No 101018682). G.N.O.L. acknowledges the financial support provided by Secretaría de Ciencia, Humanidades, Tecnología e Innovación (Secihti) through grant CBF-2025-I-201. HB acknowledges support from the European Research Council under the Horizon 2020 Framework Programme via the ERC Consolidator Grant CSF-648505. HB also acknowledges support from the Deutsche Forschungsgemeinschaft in the Collaborative Research Center (SFB 881) ``The Milky Way System" (subproject B1). YG thanks Tess Jaffe for her helpful discussion of the GB6 image from the SkyView website. This work is based on observations with the 100-m telescope of the MPIfR (Max-Planck-Institut f{\"u}r Radioastronomie) at Effelsberg. The National Radio Astronomy and Green Bank Observatory are facilities of the National Science Foundation operated under cooperative agreement by Associated Universities, Inc. This research has made use of NASA's Astrophysics Data System. This work also made use of Python libraries including Astropy\footnote{\url{https://www.astropy.org/}} \citep{2013A&A...558A..33A}, NumPy\footnote{\url{https://www.numpy.org/}} \citep{5725236}, SciPy\footnote{\url{https://www.scipy.org/}} \citep{jones2001scipy}, Matplotlib\footnote{\url{https://matplotlib.org/}} \citep{Hunter:2007}, LMFIT \citep{newville_matthew_2014_11813}, and APLpy \citep{2012ascl.soft08017R}. The continuum view of the Effelsberg survey is also available as a movie at \url{https://gongyan2444.github.io/movie/GLOSTAR_movie.m4v}.



\bibliographystyle{aa}
\bibliography{radio}

\clearpage

\begin{appendix}
\section{Supplementary data}\label{app.supplement}

\begin{figure}[!htbp]
\centering
\includegraphics[width = 0.49 \textwidth]{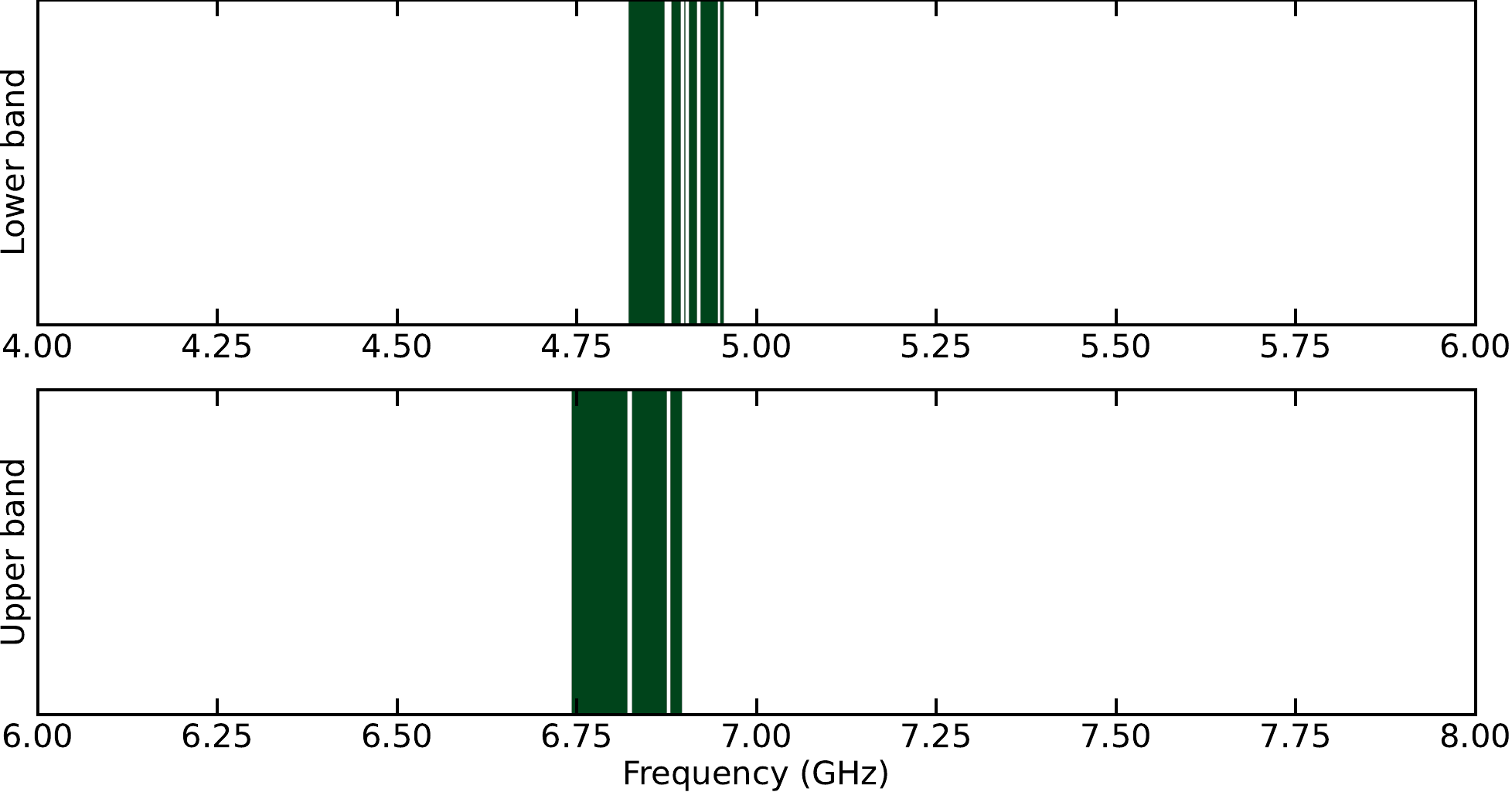}
\caption{{Waterfall plot illustrating the selected frequency channels used to construct the radio continuum images. The green-shaded regions indicate the frequency ranges included in the final continuum maps.}\label{Fig:spmask}}
\end{figure}

\begin{figure*}[!htbp]
\centering
\includegraphics[width = 0.95 \textwidth]{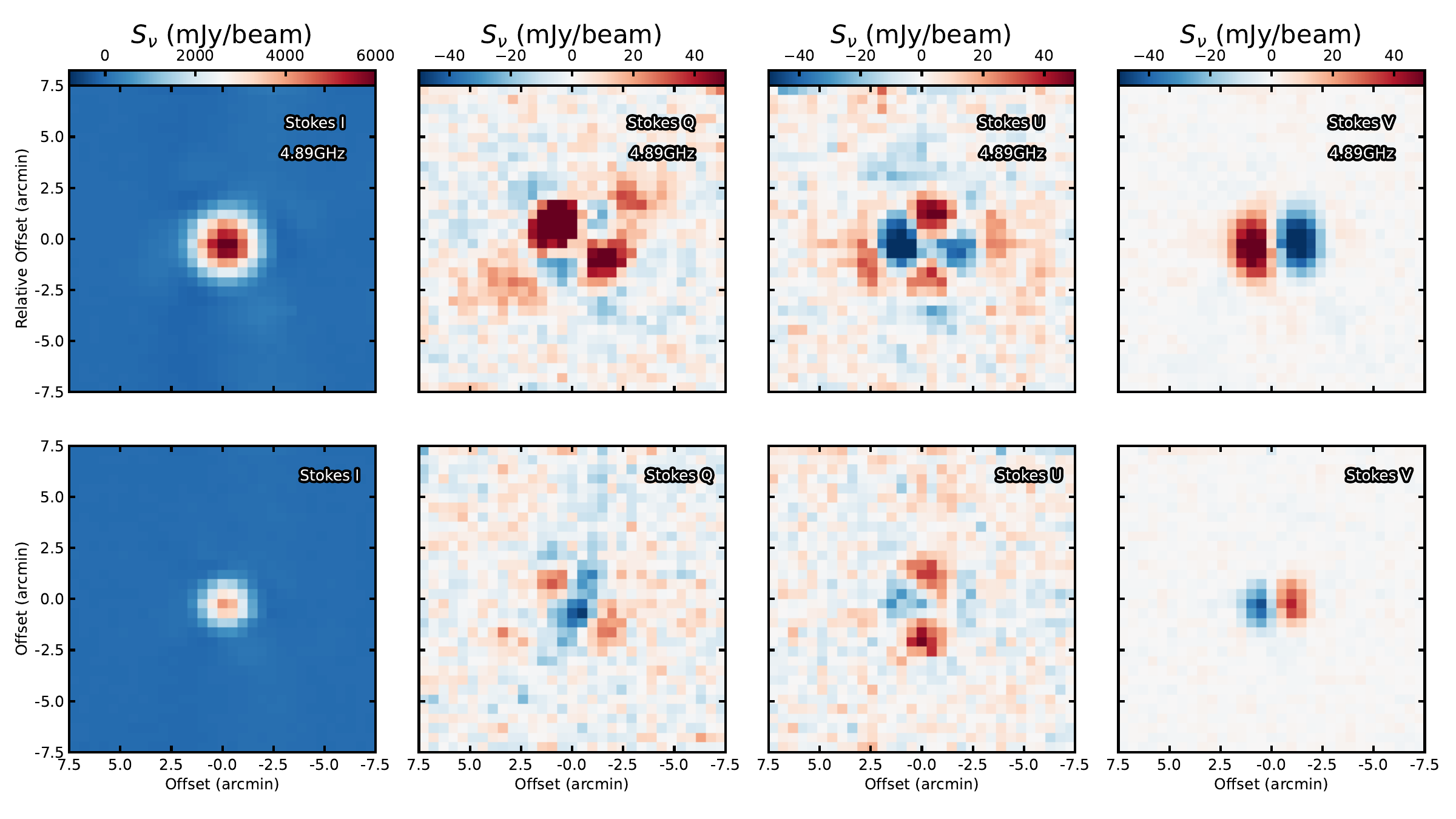}
\caption{{Stokes $I$, $Q$, $U$, and $V$ maps of 3C~295 at 4.89~GHz and 6.82~GHz. The colorbars are shown in units of mJy~beam$^{-1}$. This demonstrates a low degree of instrumental polarization.}\label{Fig:3c295}}
\end{figure*}

\begin{figure}[!htbp]
\centering
\includegraphics[width = 0.45 \textwidth]{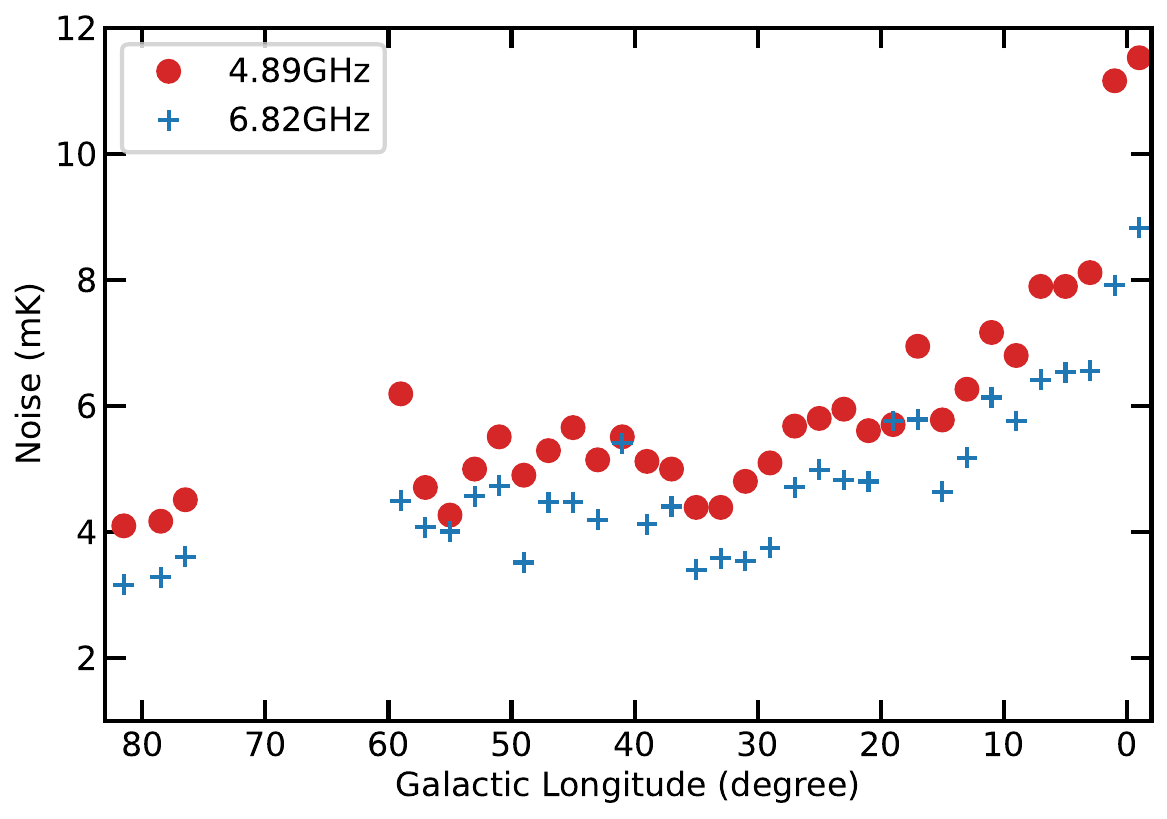}
\caption{{Total intensity noise distribution for the Effelsberg continuum surveys.}\label{Fig:noise}}
\end{figure}

\begin{figure}[!htbp]
\centering
\includegraphics[width = 0.45 \textwidth]{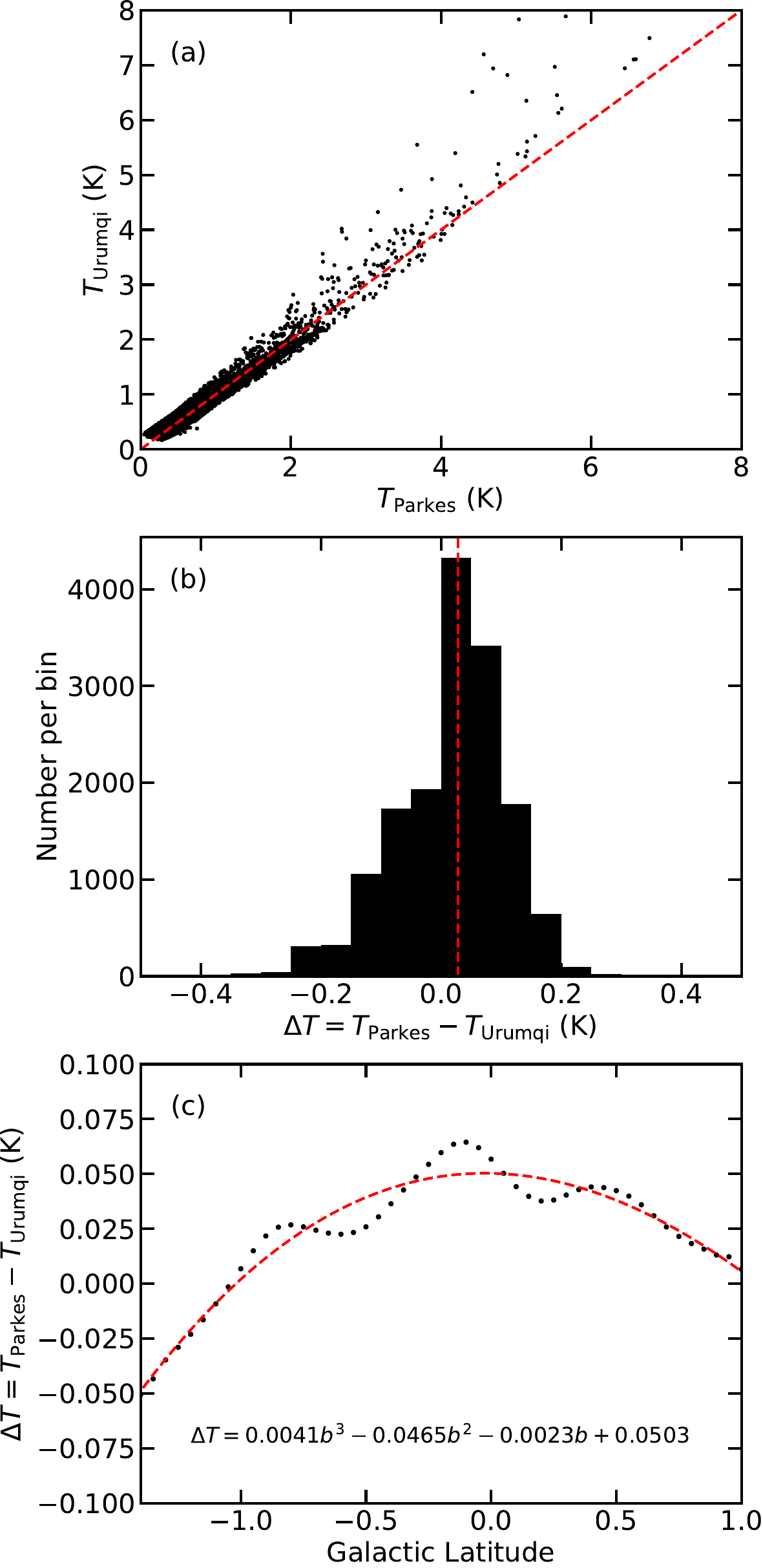}
\caption{{Comparison between the Urumqi 4.8~GHz survey and the Parkes 5~GHz survey. (a) Pixel-by-pixel comparison of the observed brightness temperatures of the two survey data sets within $10^{\circ}<\ell <30^{\circ}$. The red dashed line indicates the equality between the two data sets. (b) Histogram of the brightness temperature differences between the two surveys. The vertical dashed line denotes the median value of 0.028~K. (c) Brightness temperature difference as a function of the Galactic latitude for $\ell =11\rlap{.}^{\circ}65$. The red dashed curve represents the polynomial fit to the observed distribution.}\label{Fig:urmqi-parkes}}
\end{figure}

\begin{figure}[!htbp]
\centering
\includegraphics[width = 0.45 \textwidth]{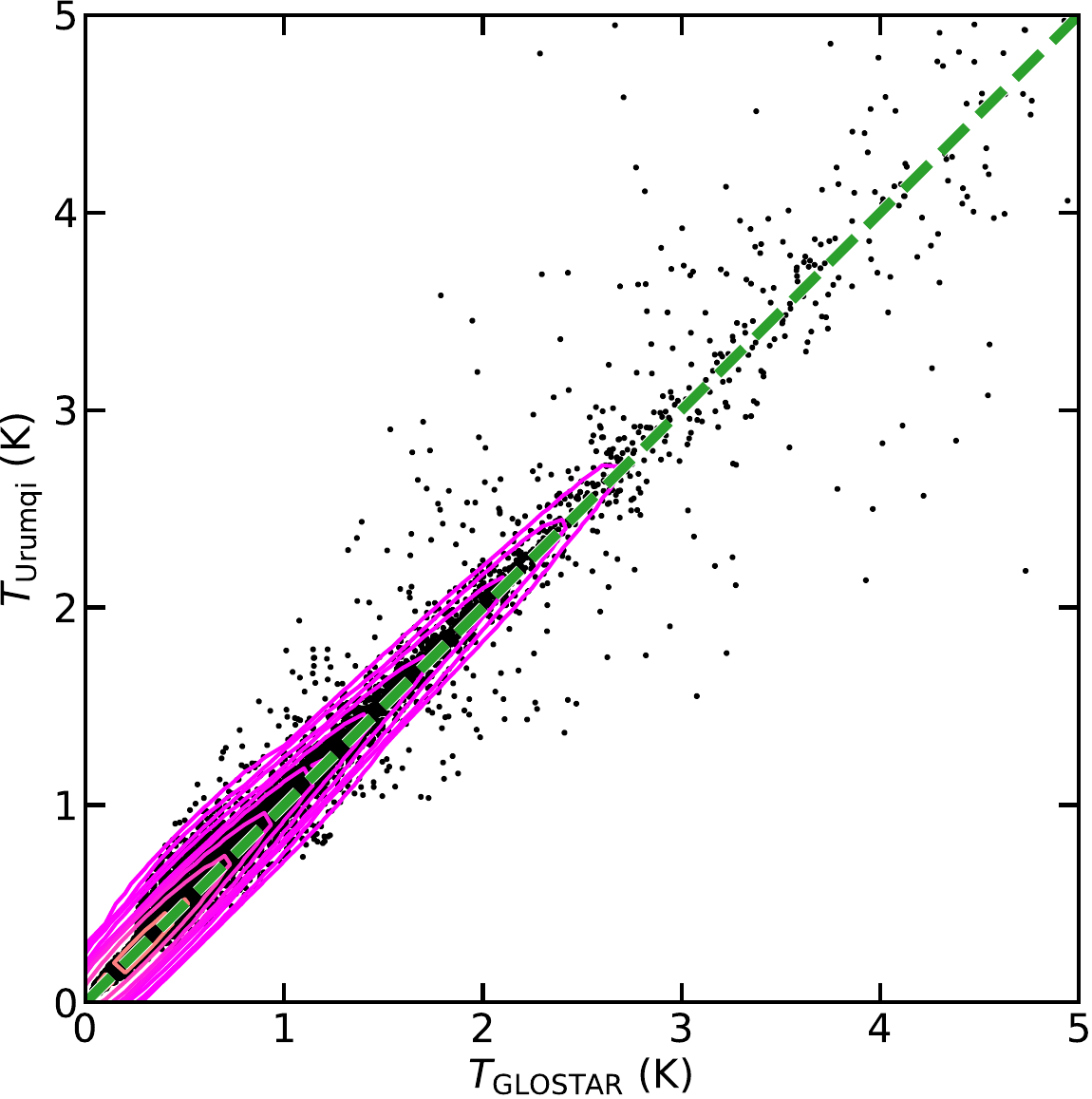}
\caption{{Pixel-by-pixel comparison of the observed brightness temperatures of the GLOSTAR and the Urumqi data within $10^{\circ}<\ell <60^{\circ}$. The purple contours, derived using Gaussian kernel density estimation, represent the point density distribution. The green dashed line indicates the equality between the two data sets.}\label{Fig:urumqi-glostar}}
\end{figure}

\begin{figure*}[!htbp]
\centering
\includegraphics[width = 1.0 \textwidth]{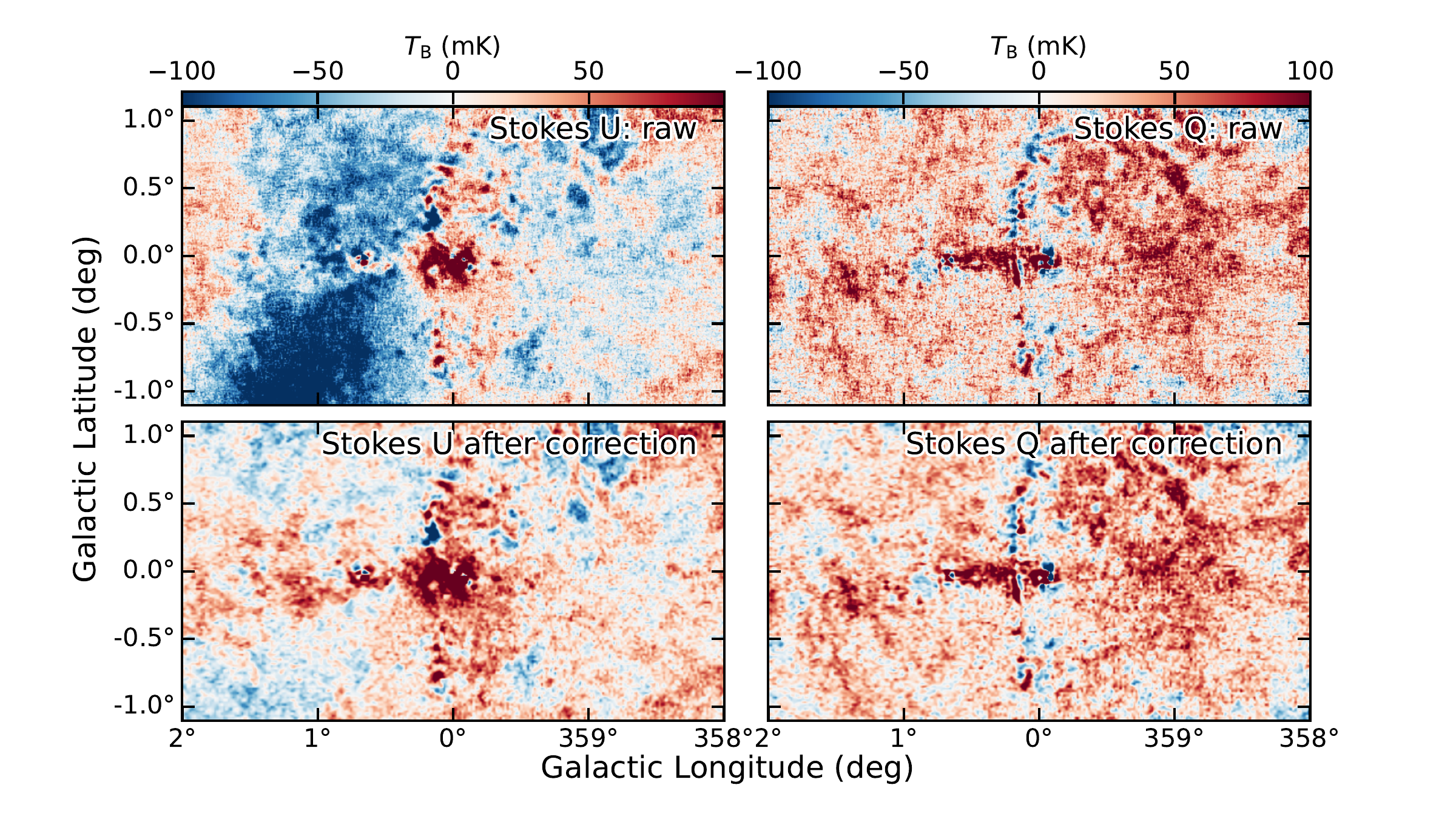}
\caption{{Zero-level restoration of the GLOSTAR 4.89~GHz polarization data in the Galactic center. The two upper panels display the GLOSTAR Stokes $Q$ and $U$ emissions in the field within $-2^{\circ}<\ell <2^{\circ}$, while the two lower panels show the corresponding Stokes $Q$ and $U$ images after correction using the WMAP polarization data.  
}\label{Fig:pol-restore}}
\end{figure*}

\begin{figure*}[!htbp]
\centering
\includegraphics[width = 1.0 \textwidth]{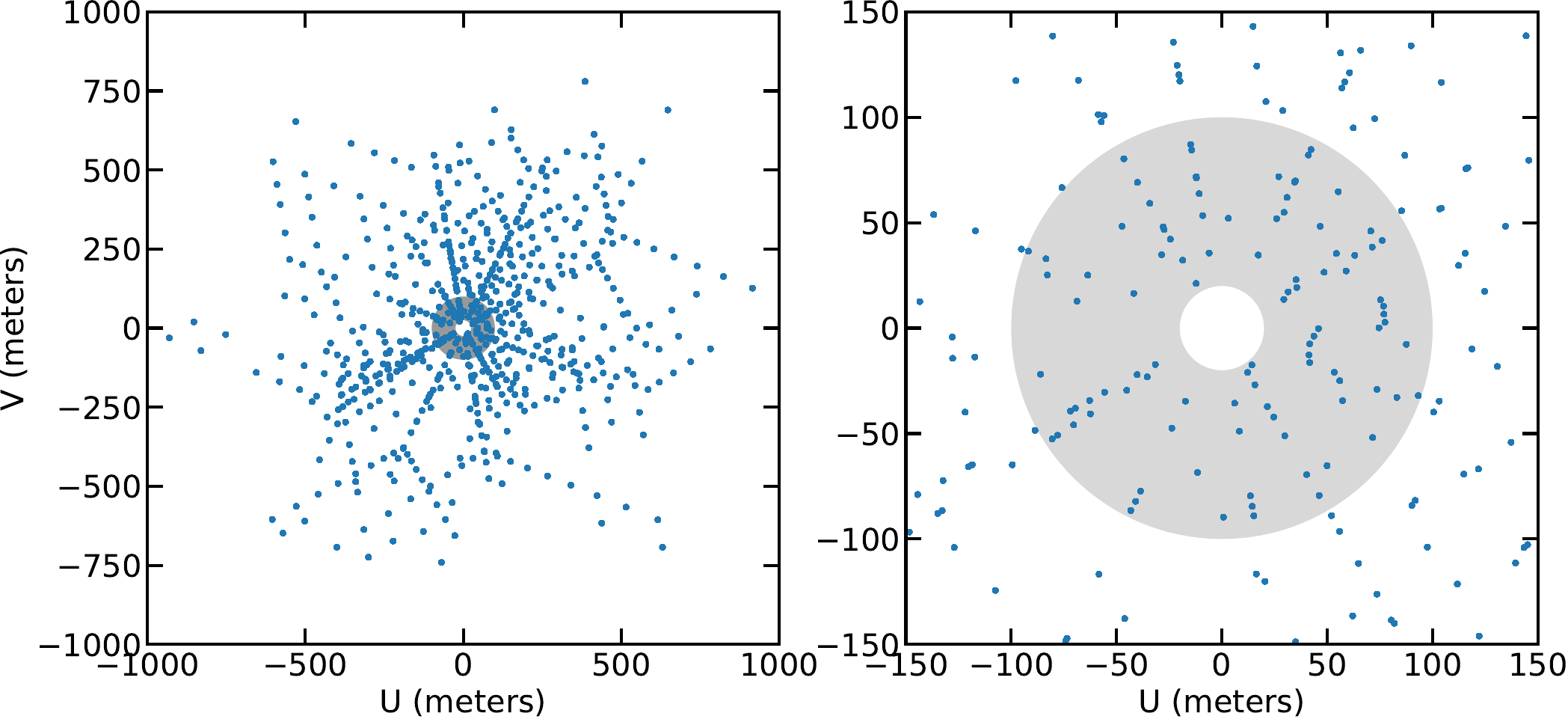}
\caption{{Left: Observed uv coverage of the VLA D-array (blue) and its overlap with the Effelsberg data (gray ring) toward a single VLA pointing. Right: Zoom in to the central part of the left panel.  
}\label{Fig:uv}}
\end{figure*}

\section{Scale Height}\label{app.height}
\begin{figure}[!htbp]
\centering
\includegraphics[width = 0.49 \textwidth]{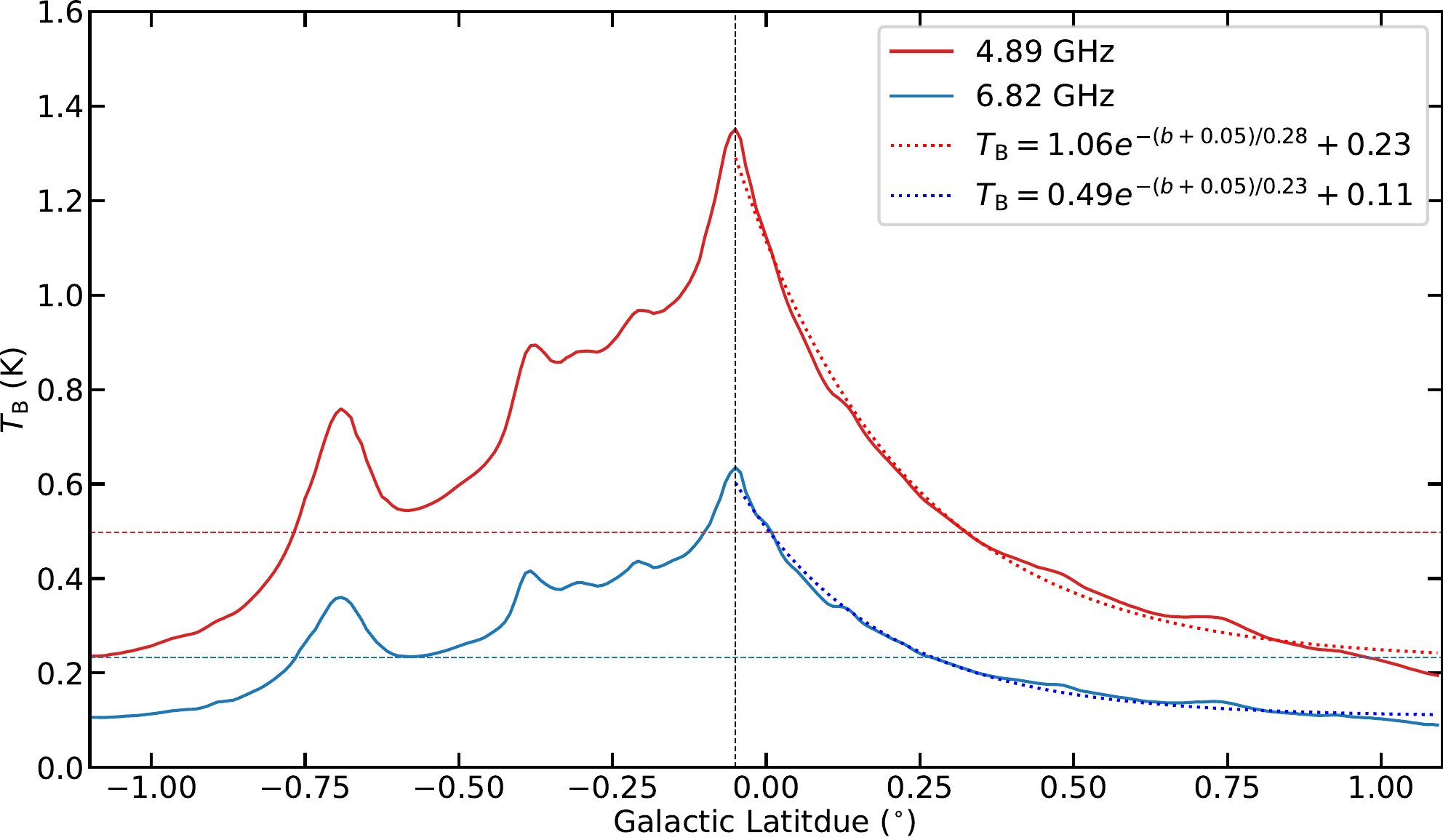}
\caption{{Mean brightness temperature profile as a function of Galactic latitude. The vertical dashed line indicates the latitude of maximum intensity. The two horizontal lines mark the $1/e$ levels of the peak intensities at the two bands. The dotted lines show the corresponding exponential fits to the profiles.
}\label{Fig:lat-profile}}
\end{figure}

Figure~\ref{Fig:lat-profile} shows the mean brightness temperature as a function of Galactic latitude, averaged over the longitude range from $-2$\degree\ to 60\degree. The resulting profile is asymmetric, with a maximum at $b=-$0\rlap{.}\degree05. The negative latitude side appears systematically brighter than the positive side, largely due to the presence of more bright radio continuum complexes at negative latitudes (see Figs.~\ref{Fig:4.89GHz-I} and \ref{Fig:6.82GHz-I}). This asymmetry is also caused by the Sun's displacement above the true Galactic mid-plane \citep[e.g.,][]{2019ApJ...885..131R}. Because the IAU coordinate system is defined relative to the Sun, objects lying in the physical mid-plane are observed at slightly negative latitudes. Adopting a Galactocentric distance of 8.2~kpc for the Sun \citep{2019ApJ...885..131R}, the observed offset of 0\rlap{.}\degree05\, corresponds to a vertical displacement of $\sim$7.2~pc toward negative Galactic latitudes, in good agreement with previous estimates of $5.5\pm5.8$~pc \citep{2019ApJ...885..131R}. 

To estimate the scale height, we only make use the profile with $b>-0.05$\degree. The distribution follows an exponential decay rather than a Gaussian distribution. Fitting it with an exponential function, we obtained $T_{\rm B}=1.06e^{-(b+0.05)/0.28}+0.23$ and $T_{\rm B}=0.49e^{-(b+0.05)/0.23}+0.11$ for 4.89~GHz and 6.82~GHz, respectively. Thus, the derived scale heights are 0\rlap{.}\degree28 and 0\rlap{.}\degree23, which corresponds to scale heights of 30--40~pc at an assumed distance of 8.2~kpc. This is significantly smaller than those (0\rlap{.}\degree4--0\rlap{.}\degree6) measured from H{\scriptsize II} regions \citep{1989ApJ...340..265W,2005AJ....130..156G}. Furthermore, the negative side appears to be have larger scale heights in terms of the $1/e$ level which can indicates scale heights of 0\rlap{.}\degree5--0\rlap{.}\degree7. As discussed earlier, more bright radio continuum complexes contribute more to the negative side. Overall, this suggests that the diffuse radio continuum emission is more tightly confined to the Galactic mid–plane than the more vertically extended population of H{\scriptsize II} regions.

\section{Large-scale template for the zero-level restoration}\label{app.template}
\begin{figure*}[!htbp]
\centering
\includegraphics[width = 0.9
\textwidth]{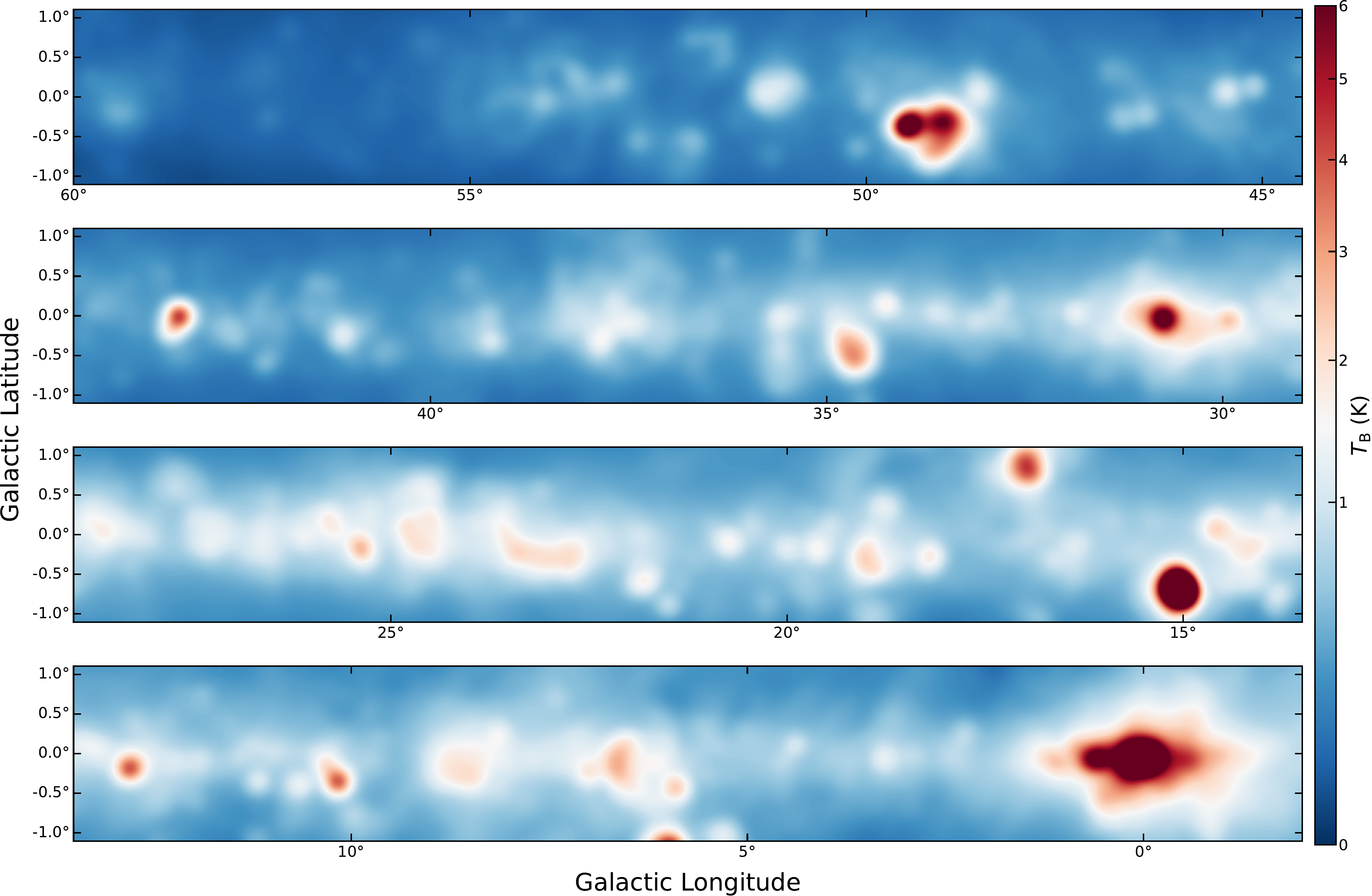}
\caption{{Distribution of the Stokes $I$ intensity from the Urumqi and Parkes combined data at 5~GHz.}\label{Fig:template}}
\end{figure*}

As discussed in Sect.~\ref{sec.zero}, we combined the Urumqi data and the corrected Parkes data into a mosaic at a common HBPW of 15\arcmin. The resulting image, shown in Fig.~\ref{Fig:template}, illustrates a smooth transition from the Urumqi data to the Parkes data, offering a template for the large-scale distribution used in the zero-level restoration of our GLOSTAR data. This template was further used to correct the zero-level offsets, and the distributions of the restored Stokes $I$ maps are shown in Figs.~\ref{Fig:4.89GHz-I} and \ref{Fig:6.82GHz-I}.

\end{appendix}

\end{document}